\documentclass[sigconf,table,balance=false]{acmart}

\makeatletter
\def\@ACM@checkaffil{
    \if@ACM@instpresent\else
    \ClassWarningNoLine{\@classname}{No institution present for an affiliation}%
    \fi
    \if@ACM@citypresent\else
    \ClassWarningNoLine{\@classname}{No city present for an affiliation}%
    \fi
    \if@ACM@countrypresent\else
        \ClassWarningNoLine{\@classname}{No country present for an affiliation}%
    \fi
}
\makeatother

\usepackage{tikz}
\usepackage{amsmath}
\usepackage{float}
\usepackage{filecontents}
\usepackage{tabularx}
\usepackage{booktabs}
\usepackage{graphicx}
\usepackage{hyperref}
\usepackage{hhline}
\usepackage{subcaption}
\usepackage{enumitem}

\usepackage{float} 
\usepackage{soul,color}
\usepackage[utf8]{inputenc}
\usepackage{longtable}
\usepackage[flushleft]{threeparttable}
\usepackage{chngcntr}
\usepackage{xcolor}
\definecolor{editCol}{rgb}{0.0, 0.0, 0.0}
\newcommand{\edit}[1]{{\textcolor{editCol}{#1}}}
\usepackage{pbox}
\usepackage{multirow}

\usepackage{array}
\usepackage{longtable}
\usepackage{makecell}
\usepackage{afterpage} 
\usepackage[title]{appendix}
\usepackage{subcaption,graphicx}
\usepackage{ragged2e}

\newcolumntype{L}[1]{>{\raggedright\let\newline\\\arraybackslash\hspace{0pt}}m{#1}}
\newcolumntype{C}[1]{>{\centering\let\newline\\\arraybackslash\hspace{0pt}}m{#1}}
\newcolumntype{R}[1]{>{\raggedleft\let\newline\\\arraybackslash\hspace{0pt}}m{#1}}

\usepackage{popets}
	\setcopyright{popets}
	\copyrightyear{YYYY}
	\acmYear{YYYY}
	\acmVolume{YYYY}
	\acmNumber{X}
	\acmDOI{XXXXXXX.XXXXXXX}
	\acmISBN{}
	\acmConference{Proceedings on Privacy Enhancing Technologies}
\begin{document}

\title[Examining Caregiving Roles to Differentiate the Effects of Community Oversight]{Examining Caregiving Roles to Differentiate the Effects of Using a Mobile App for Community Oversight for Privacy and Security}

\author{Mamtaj Akter}
\orcid{0000-0002-5692-9252}
\affiliation{%
  \institution{New York Institute of Technology, USA}
  \city{}
  \state{}
  \country{}
}

\author{Jess Kropczynski}
\orcid{0000-0002-7458-6003}
\affiliation{%
  \institution{University of Cincinnati, USA}
  \city{}
  \state{}
  \country{}
}

\author{Heather Lipford}
\orcid{0000-0002-5261-0148}
\affiliation{%
  \institution{University of North Carolina, Charlotte, USA}
  \city{}
  \state{}
  \country{}
}

\author{Pamela Wisniewski}
\orcid{0000-0002-6223-1029}
\affiliation{
  \institution{Vanderbilt University, USA}
  \city{}
  \state{}
  \country{}
}

\renewcommand{\shortauthors}{Akter et al.}

\begin{abstract}
We conducted a 4-week field study with 101 smartphone users who self-organized into 22 small groups of family, friends, and neighbors to use ``CO-oPS,'' a mobile app for co-managing mobile privacy and security. We differentiated between those who provided oversight (i.e., caregivers) and those who did not (i.e., caregivees) to examine differential effects on their experiences and behaviors while using CO-oPS. Caregivers reported higher power use, community trust, belonging, collective efficacy, and self-efficacy than caregivees. Both groups' self-efficacy and collective efficacy for mobile privacy and security increased after using CO-oPS. However, this increase was significantly stronger for caregivees. 
Our research demonstrates how community-based approaches can benefit people who need additional help managing their digital privacy and security. We provide recommendations to support community-based oversight for managing privacy and security within communities of different roles and skills.
\end{abstract}

\keywords{Community; Community Oversight; Collaborative; Permission; Mobile; Apps; Mobile Privacy; security; Privacy; Caregiver; Efficacy}

\maketitle

\section{Introduction}

A recent Pew Research study \cite{vogels_americans_2019} reported that most adults in the U.S. have significant knowledge gaps regarding their digital privacy and security. The report suggests a lack of understanding of how third party entities access personal information online \cite{auxier_americans_2019}, making users susceptible to privacy and security breaches. This begs the question of how we might close this knowledge gap? \edit{Previous studies have explored solutions involving one party overseeing another, such as organizations using Mobile Device Management systems to secure employee data \cite{hayes_effective_2020} and parents monitoring teens through mobile online safety apps \cite{wisniewski_parental_2017}. These studies suggested that collaborative approaches, rather than unilateral controls, can offer better privacy protection by fostering collective learning and discussion ~\cite{akter_from_2022}. Crowdsourcing has also been proposed as a method to improve individual's mobile privacy decisions such as managing app permissions to safeguard personal information  \cite{Lin_expectation_2012, ismail2017permit, zhang_privacy_2016, liu_understanding_2019}, though it often lacks reliability \cite{rashidi_Android_2018}.} An approach that has shown promise is to leverage the social influence of informal networks to infuse expertise and exert influence on privacy and security decisions \cite{mendel_susceptibility_2017}. For example, several studies have shown that people tend to trust and follow privacy advice from their trusted circles, often turning to friends and family for guidance on digital privacy and security topics \cite{mendel_susceptibility_2017, das_effect_2014, das_increasing_2014}. 

Therefore, networked privacy researchers have called for community based approaches to co-manage digital privacy and security, where individuals in a trusted community can help one another equally make individual and collective privacy and security decisions \cite{wan_appmod_2020, chouhan_co-designing_2019, aljallad_designing_2019}. \edit{While this body of research confirms that people often rely on their social connections to inform their digital privacy and security decisions, it has yet to explore how these dynamics shift when relationships involve knowledge imbalances or power hierarchies, such as caring responsibilities for children or older adults.} Our prior studies \cite{akter_from_2022, akter_evaluating_2023, chouhan_co-designing_2019, kropczynski_examining_2021} have shown how communities often consist of hierarchical (i.e. uneven) levels of expertise and knowledge.  Individuals with more expertise often take the role of ``caregivers'' to provide support and guidance to others who are less tech-savvy \cite{van_parys_you_2019, correa_brokering_2015, kiesler_troubles_2000, poole_computer_2009}. Other individuals in the community are at the receiving end of the privacy and security support \cite{kropczynski_towards_2021, anaraky_disclose_2021}, and are thus considered ``caregivees". However, community-based collaborative approaches may not take into account these unequal levels of expertise, as they have been primarily designed to support community members equally, with individuals giving and receiving feedback. Recently, we \cite{akter_evaluating_2023} conducted a field study using a community oversight app that allowed individuals in a community to help each other equally and found that users, in general, benefited from using the app. 
Inspired by these positive results, we are now motivated to further investigate the impact of unequal roles within these collaborative environments.


This paper aims to address this gap, \textcolor{black}{by investigating: 1) how caregivees vs. caregivers differ, 2) how mechanisms designed for collaborative community-based privacy and security support may affect caregivers and caregivees differently, and 3) if there are distinct differences in privacy behaviors between these two groups.} \textcolor{black}{We} developed and deployed a community oversight app, called "CO-oPS" \cite{akterCOoPS2023} that allows people of a trusted community to review one another's mobile privacy and security decisions and provide advice and feedback on them. Our primary goal in this paper is to investigate how differently caregivers, who provide oversight, and caregivees, who receive the oversight, would be impacted by using the app. \textcolor{black}{By providing insights into the varying effects of the CO-oPS app on these two unequal roles, we expect to gain generalizable knowledge that helps researchers and designers better understand the phenomena of privacy and security caregiving, how privacy-based interventions may effect these groups differently, and how to redesign future privacy-based interventions tailored to the unique needs of caregivers and caregivees.} As such, we seek to answer the following high-level research questions:
\begin{itemize}
\item \textbf{RQ1:} \textit{How do the characteristics of Caregivers versus Caregivees differ?} 
\item \textbf{RQ2:} \textit{How do the perceived outcomes of using the CO-oPS app differ based on Caregiving status}?
\item \textbf{RQ3:} \textit{How do the privacy behaviors of Caregivers \edit{and} Caregivees differ when using the CO-oPS app?}
\end{itemize}
To answer these research questions, we conducted a field study with 22 self-organized groups of 2-6 smartphone users. First, participants within each group reported their pre-existing caregiving relationships by identifying the individual(s) in their community for whom they go to for advice and questions regarding their mobile privacy and security, whom we labeled as caregivers. The rest of the community members were labeled as caregivees. We then had our communities of caregivers and caregivees use the CO-oPS app for four weeks, complete pre-post-study surveys, and participate in optional follow-up interviews. To answer RQ1, we first examined how the caregivers and caregivees are different in terms of their demographic information and power usage  \edit{- individual's propensity to maximize the use of technology to its fullest} \cite{sundar2010personalization}. For RQ2, we measured how our caregivers' and caregivees' perceptions for community trust \cite{chouhan_co-designing_2019}, community belonging \cite{carroll_community_2003, kropczynski_examining_2021}, community collective efficacy \cite{carroll_collective_2005, kropczynski_examining_2021}, and self-efficacy \cite{bandura1982self, kropczynski_examining_2021} differed and changed after using the CO-oPS app. For RQ3, we measured the between-group differences among the pairs of caregivers and caregivees' app activities they performed using the CO-oPS app. We also qualitatively analyzed their interview transcripts to further investigate how their perceptions towards the use of the CO-oPS app differed when they provided or received caregiving. 


Overall, we found that caregivers reported higher power usage compared to caregivees, with no significant differences based on demographic characteristics (RQ1). For RQ2, caregivers showed higher levels across all pre-study measures—community trust, community belonging, community collective efficacy, and self-efficacy—compared to caregivees. Both groups experienced significant increases in self-efficacy and community collective efficacy in managing mobile privacy and security after using the CO-oPS app, with caregivees showing a significantly stronger increase. For RQ3, tensions arose around the privacy feature of the CO-oPS app, as caregivees expressed concerns about app usage privacy and thus hid a significantly greater number of installed apps, while caregivers viewed this feature as counterproductive to community oversight.

In summary, our study highlights the importance of differentiating caregiving roles in communities for collaborative mobile privacy and security management. It uniquely contributes to networked privacy research by examining collaborative mobile privacy and security management in the context of caregiving in communities to support individuals who need additional help. Specifically, our research: 1) Untangles the behavioral and perceptual differences between caregivers and caregivees when using a community oversight app; 2) Provides empirical evidence of the potential for community oversight to improve caregivees' decisions on mobile privacy and security; and 3) Offers design recommendations for features aimed at supporting communities in providing and receiving caregiving tailored to the specific needs of individuals.

\section{Background}
\textbf{Community-based Approaches for Privacy and Security}\\
A Pew Research study ~\cite{noauthor_majority_2015} showed that 77\% of U.S. smartphone users have reported downloading different third party mobile apps on their phones. These third party apps access users' personal and sensitive information ~\cite{nw_mobile_2015}, often without users' informed consent, as Reardon et al. reported in \cite{reardon_50_2019}. However, another Pew research study reported that a majority of U.S. adults have significant knowledge gaps and lack understanding of how to control and manage their digital privacy and security \cite{vogels_americans_2019}. Users even find it difficult to understand the language of app permission requests that third party apps prompt before accessing users' information ~\cite{ferreira_securacy_2015} and, therefore, they struggle to decide whether to grant an app permission or not \cite{kelley_conundrum_2012}. To help individuals with their digital privacy and security, 
privacy researchers have examined crowdsource-based approaches \cite{zhang_privacy_2016, liu_when_2018, Lin_expectation_2012, liu_understanding_2019} that leveraged the knowledge and information of other users to provide recommendations to users so that they can make informed decisions about their app permissions. For example, Zhang et al. \cite{zhang_privacy_2016} proposed a social nudge mechanism that used the idea of \textit{visibility of social norms} and allowed users to make app permission decisions based on the percentage of other users who approved different data permissions for each installed app. Through an online experiment, they evaluated this interface and found that users mostly perceived this social nudge as a social norm or collective expectation of information sharing, and users tended to approve their apps’ data use when they saw that more users accepted those apps' permissions. 

While these crowdsource-based solutions help individuals by raising their awareness about other users' mobile privacy practices, they give little consideration to the trustworthiness of the information. Individuals might benefit more if the information comes from expert users who have expertise in mobile app privacy and security. Hence, Rashidi et al. \cite{rashidi_Android_2018} incorporated this idea of "expert user" into their permission control framework titled "DroidNet" which provided users recommendations based on the decisions made by the majority of expert users in the network. 
However, a large body of research work also demonstrated that individuals learn and change their digital privacy and security behaviors when they become aware of their close trusted circle's privacy and security practices. For example, Das et al.'s studies \cite{das_increasing_2014, das_role_2015} demonstrated the effectiveness of "social proof" - i.e. being able to view how many friends in their social network use a specific security feature - and showed that individuals are more influenced to adopt a privacy and security feature when they witness adoption by others. Redmiles et al. also found that people trust and adopt privacy advice when it comes from their trusted close circles, e.g., family members, friends, and co-workers \cite{redmiles2016think}. In recent work, Mendel and Toch \cite{mendel_social_2023} conducted a field experiment to evaluate a social support mechanism and found that users relied significantly more on the advice of close connections to learn about mobile security than community volunteers. 
Identifying the importance of social support in digital privacy and security, networked privacy researchers have investigated collaborative approaches. For instance, Wan et al. \cite{wan_appmod_2020} designed a mobile app, titled AppMoD, to help individuals delegate their mobile privacy and security questions to an advisor - someone in their trusted social network. The advisor had the ability to make app permission decisions on behalf of the user. In our prior work \cite{chouhan_co-designing_2019}, we proposed a community oversight framework to allow individuals in a trusted community, e.g., friends, family members, and co-workers, to help each other make their mobile privacy and security decisions together as a team. \\

\noindent
\textbf{Caregiving for Security and Privacy}\\
Communities can be heterogeneous when it comes to knowledge, behavior, skills, and roles in mobile privacy and security. Previous work examined how individuals in a community can be different in privacy and security knowledge, as well as take different roles in providing support. For example, our prior studies \cite{akter_from_2022, akter_it_2023, akter2024towards} revealed that although parents tended to take the role of supervising their families' mobile online safety, privacy, and security, tech-savvy teens showed more concerns about the permissions granted on their parents' phones, while parents were more focused on the kinds of apps their teens used. Researchers also examined how different age groups in communities vary in their decision-making about digital privacy and security. For example, Huang and Bashir  \cite{huang2018surfing} identified that older adults, e.g., 65 years or older, tend to take fewer privacy protections than their younger peers, e.g., 55 to 64 years. However, in a more recent study, Anaraky et al. \cite{anaraky_disclose_2021} compared privacy decision making of young adults and older adults and found that young adults were more likely to disclose information based on their trust in the apps, while older adults, being careful, considered the pros and cons of each information disclosure. This body of research demonstrates that in communities, there may be some individuals who are more knowledgeable in making mobile privacy decisions and/or providing support to others (e.g., caregivers), and some might lack the knowledge and seek support from others (e.g., caregivees). We have previously explored this concept of caregiving in the context of providing and receiving technological support within close, trusted circles, such as friends and family \cite{kropczynski_examining_2021}. This study extends this concept to examine caregiving in the management of mobile privacy and security. 

The core idea behind collaborative community-based approaches \cite{wan_appmod_2020, chouhan_co-designing_2019, aljallad_designing_2019} is to empower individuals within trusted communities to practice social support \cite{dourish_social_2005, Das_social_2016}, enabling them to collaboratively assist one another in making both individual and collective privacy and security decisions. However, one of our prior work \cite{akter_evaluating_2023} suggested that these mechanisms are most effective when there are disparities in knowledge and expertise within the community, allowing those in need of support with mobile privacy and security decisions to benefit from the guidance of more knowledgeable community members. \edit{Despite this, the collaborative community-based mechanisms in the existing literature have not adequately considered the imbalance of expertise, caregiving roles, and the varying needs and perspectives \cite{badillo2020towards, park_towards_2023} that accompany them. Therefore, the objective of this paper is to investigate how these approaches, which treat all community members as equals, may or may not benefit caregivees who require additional support, as well as how they help caregivers support others.} To examine these, we developed a collaborative community-based mobile privacy and security management app, CO-oPS, and conducted a study involving 22 groups of caregivers and caregivees who used the app within their groups. In the following section, we present a design overview of the CO-oPS app.


\begin{figure}[t]
\centering
\begin{subfigure}[t]{.236\textwidth}\centering
  \includegraphics[width=\columnwidth]{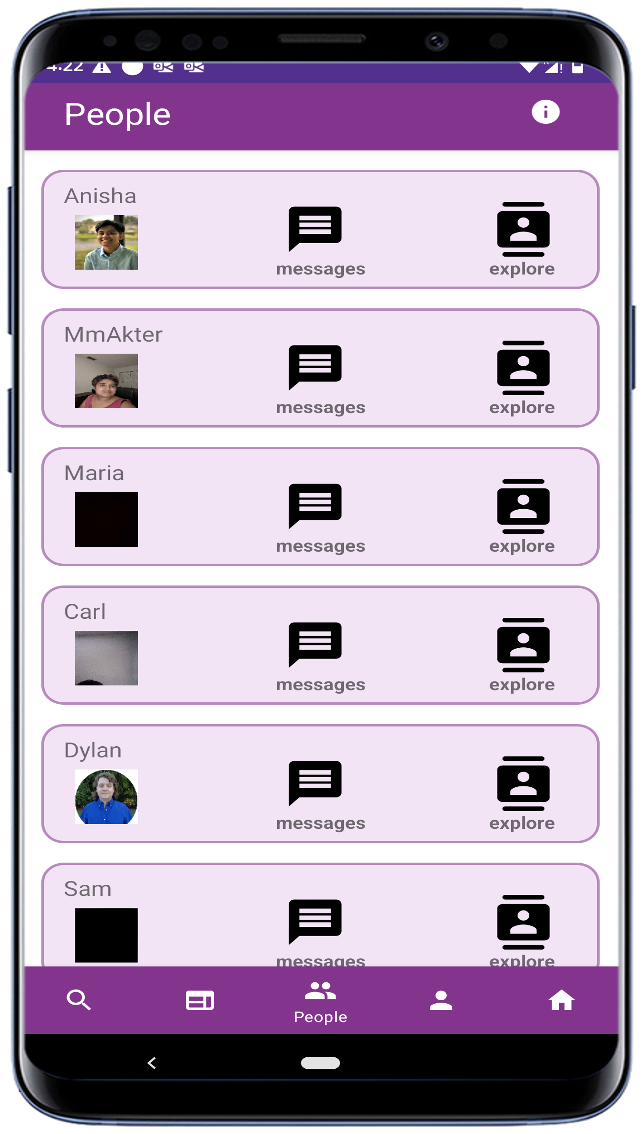}
  \caption{}
\end{subfigure}%
\begin{subfigure}[t]{.238\textwidth}\centering
  \includegraphics[width=\columnwidth]{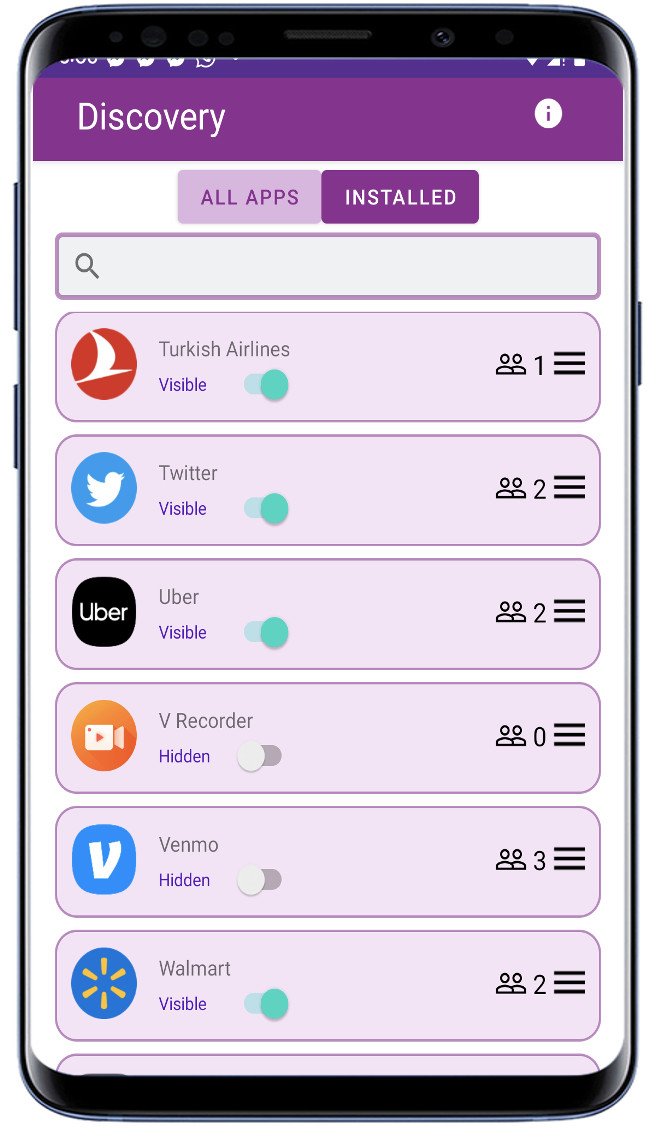}
  \caption{}
\end{subfigure}
\begin{subfigure}[t]{.236\textwidth}\centering
  \includegraphics[width=\columnwidth]{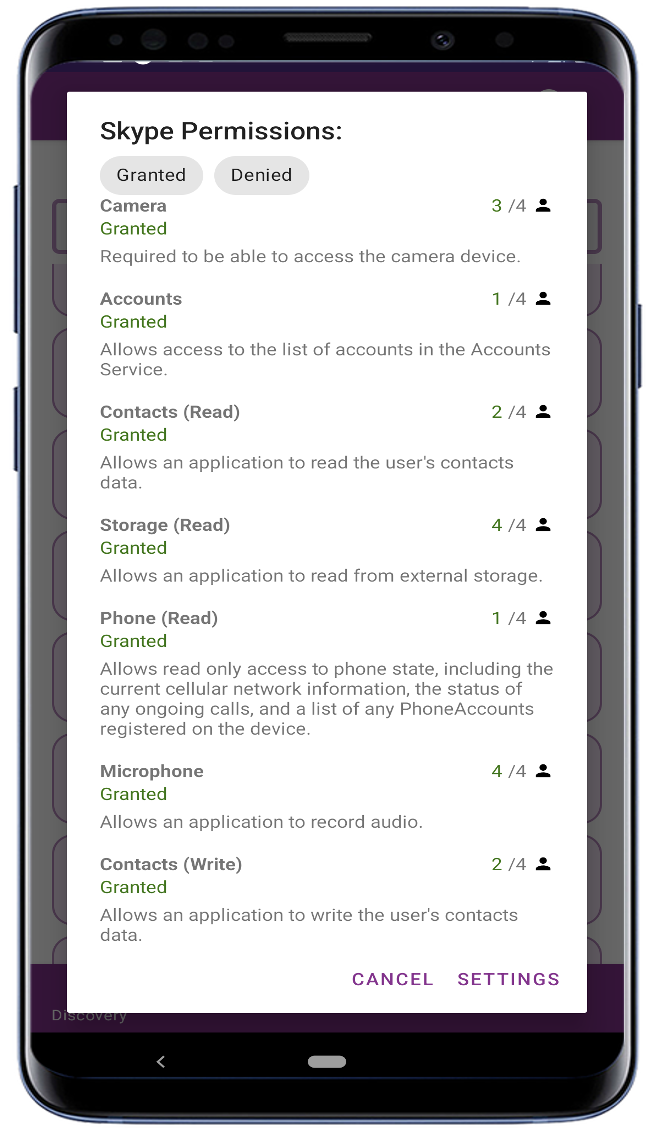}
  \caption{}
\end{subfigure}
\begin{subfigure}[t]{.236\textwidth}\centering
  \includegraphics[width=\columnwidth]{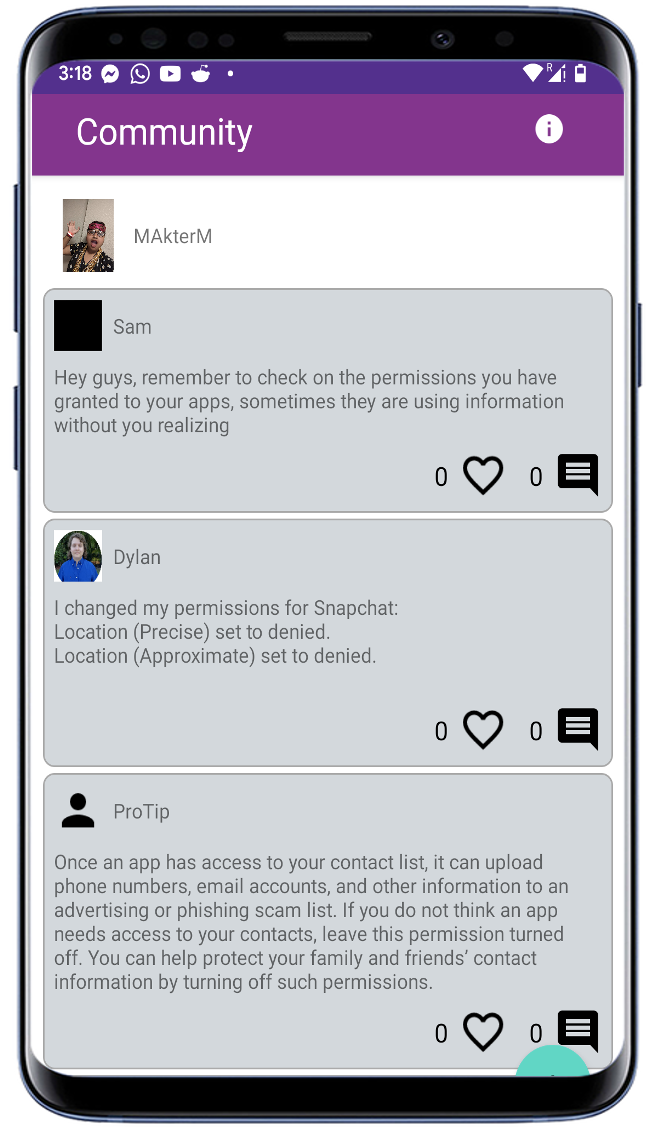}
  \caption{}
\end{subfigure}
\caption{CO-oPS App Interfaces: (a) People, (b) Discovery, (c) Permissions, and (d) Community Feed.}~\label{fig:app}
\end{figure}
\section{Design of the CO-oPS App}
We developed the Android CO-oPS app \cite{akterCOoPS2023} based on our proposed model of community oversight for privacy and security proposed by Chouhan et al. \cite{chouhan_co-designing_2019}. This app allows individuals in a trusted community, e.g., friends, families, and co-workers, to review one another's apps installed and permissions granted, and provide or seek advice from one another, providing all community members the same interface and functionalities. Our app includes three main categories of features: 1) \edit{Co-monitoring}, 2) Communication, and 3) Privacy.  

\textit{\textbf{\edit{Co-monitoring Features: }}}These features allow users to view the list of community members (People page, \autoref{fig:app}a). Here, they can explore the apps installed on each member's device through the Discovery page (\autoref{fig:app}b). For any installed app within the community, users have the ability to review the permissions granted or denied to that specific app on the Permission page (\autoref{fig:app}c). This feature enables community members to assess each other's choices regarding mobile privacy and security, facilitating joint learning and feedback. The Permission page displays a comprehensive list of permissions granted or denied to a particular app, alongside the number of community members who have made the same choices. Additionally, it offers users a convenient "SETTINGS" link, redirecting them to Android Settings, where they can adjust the permissions of their own installed apps.

\textit{\textbf{Communication Features:}} The communication features keep users connected regarding one another's decisions, questions, and feedback. Here, users can stay updated on the permission changes made in the community, receive weekly tips, and engage in open discussion or private messaging. For instance, when permission changes are made using the above "SETTINGS" link, the app automatically generates a post on the Community Feed, as shown in \autoref{fig:app}d, to inform the entire community. The Community feed also fosters open discussion among community members. Moreover, this feed provides weekly tips to educate users on safe app usage and permissions. For example, one of the weekly tips was ``Pay careful attention to the app permission prompts. If the permission is not required for the app to function, consider denying." Lastly, the People page (\autoref{fig:app}a) allows one-to-one messaging for asking questions or offering guidance.

\textit{\textbf{Privacy Feature:}} This feature provides users with the option to control the visibility of their installed apps (Discovery page, \autoref{fig:app}b), ensuring personal privacy regarding their app usage within the community. By allowing users to hide specific apps, they can selectively choose which ones are visible to others. When an app is hidden on a community member's device, it becomes inaccessible for review by other community members on the Discovery page. Moreover, the changes made to the permissions of these hidden apps are not posted on the community feed page, further safeguarding the privacy of users' app-related activities.

\section{Methods}
The overall goal of this paper is to examine the key differences in privacy and security behaviors between the individuals who provide oversight and those who receive when using a community oversight app. To achieve these goals, we recruited 22 small self-organized groups of people (2 to 6 individuals), who knew each other. \edit{In total, 101 participants were recruited across these groups, starting with primary contacts who then invited others within their social circles to join. } Participants installed and used the CO-oPS Android app for four weeks with their group to exchange support and guidance on their privacy decisions. Participants also completed pre- and post-study surveys and were invited to participate in optional follow-up interviews. \edit{The study received approval from the Institutional Review Boards (IRBs) of the universities involved.} \\

\noindent
\textbf{Study Procedures}\\
\textbf{\textit{Pre and Post Study Surveys}:} Our participants completed a Qualtrics survey before and after the four-week long field study. In the pre-study survey, we first asked participants to describe the interpersonal relationships (e.g., friend, family member, neighbors, etc.) they had with each member of their community, and the proximity to the others' residences with respect to their own. To distinguish the caregivers and caregivees within the groups, we asked our participants about their pre-existing mobile privacy and security caregiving relationships. To this end, we asked them to report who in their community they go to for questions regarding their mobile privacy and security decisions, e.g., whether an app is safe to install or a permission is safe to allow for an app. \edit{We did not inform our participants that we would differentiate them as caregivers and caregivees as we wanted to avoid potential conflicts within their groups and to minimize bias in their app interactions and mobile privacy behaviors.} Next, we collected participants' demographic information, i.e., age, gender, ethnicity and education level. In both pre- and post-study surveys, we also measured some prevalidated and newly created constructs.

\begin{table}[]
 \footnotesize
 \caption{Weekly App Tasks}
  \label{tab:app-tasks}
\begin{tabular}{|p{8cm}|}
\hline
\textbf{Week-1:} 
1. Review your own apps from the “Discovery” page > “Installed” tab. Hide the apps that you do not want others to see.  
2. Review your community’s apps from the Discovery > “All Apps”. Check if you have any uncommon apps that no one else is using.  
3. For the apps you have in common with others, compare the permissions you granted but others denied.\\ \hline

\textbf{Week-2:} 1. Read the weekly pro tip and add a comment there. 2. From the “People” page, review one of your community member’s apps. Check if there are any apps or permissions that may not be safe. 3. Send a message to warn them about unsafe apps or permissions. \\ \hline
  
\textbf{Week-3:} 1. Read the weekly pro tip and add a comment there.  \newline 2. Review your own apps and permissions and check if any granted permissions may be unsafe. Consider changing those permissions. 3. Review the apps and permissions of someone in your community. Let them know if they have any unsafe permissions.\\ \hline
  
\textbf{Week-4:} 1. Check the messages received from your community. Consider changing the apps and permissions accordingly. 2. Review your community members’ apps and check if any unsafe apps or permissions exist. 3. Write a post on the Community feed to warn others about the unsafe apps or permissions found. \\ \hline
\end{tabular}
\end{table}

\textbf{\textit{Weekly App Tasks:}} At the beginning of the field study, community members were asked to install the CO-oPS app on their phones and setup their user profiles. \edit{A key step in this setup process involved the app launching the Discovery page (\autoref{fig:app}b) with the prompt: "Select the apps that you want to hide from your community," allowing participants to choose which of their apps would be hidden from others. This ensured that participants could hide their installed apps before they became visible to their group members.} \textcolor{black}{To this end, we allowed our participants to decide the apps they wanted to hide from their group as we did not want to bias their choices.} Each week, our participants received a list of tasks on the home page of the CO-oPS app, where they could check off each task after completion.  However, we did not follow up with participants for not completing their tasks, as we wanted them to use the CO-oPS app as they saw fit to provide or receive privacy caregiving from one another. The weekly tasks are presented in \autoref{tab:app-tasks}. Upon completing the week-4 tasks and post study survey, participants were each compensated with a \$40 Amazon gift card. N=3 participants discontinued after the second week and were compensated with a \$20 Amazon gift card. However, we discarded all data collected from those participants who stopped participating.

\textbf{\textit{Follow-up Interviews: }} At the end of the field study, we invited all participants to optional one-on-one Zoom interviews which allowed us to further explore how they perceived the CO-oPS app features for providing or receiving privacy and security support to/from their communities. N=51 participants volunteered for follow-up interviews, each of whom were compensated with a \$10 Amazon gift card. Appendix-B presents the sample interview questions we asked during the interviews. The interview sessions took 40 to 70 minutes to complete, and were audio/video recorded. \\

\begin{table*}[t]
 \footnotesize
 \centering
  \caption{Hypotheses Summary}
  \label{tab:all_hypotheses}
\begin{tabular}{llll} 
\hline
\textbf{Constructs}  & \textbf{Main Effect of Caregiving} & \textbf{Main Effect of CO-oPS} & \textbf{Interaction Effect}  \\ 
  \hline
\addlinespace[1ex]
 \textbf{H1:} PU* &  
Caregiver$_{PU}$>Caregivee$_{PU}$
&  -- 
&  --  \\
 \addlinespace[1ex]
 \textbf{H2:} CT* 
&  a) Caregiver$_{CT}$<Caregivee$_{CT}$ 
&  b) PreStudy$_{CT}$<PostStudy$_{CT}$
&  c) Caregiving x PrePost|Caregivee$\rightarrow$(+)CT \\
 \addlinespace[1ex]
 \textbf{H3:} CB* 
&  a) Caregiver$_{CB}$=Caregivee$_{CB}$
&  b) PreStudy$_{CB}$ < PostStudy$_{CB}$
&  c) Caregiving x PrePost|Caregivee$\rightarrow$(+)CB   \\
 \addlinespace[1ex]
 \textbf{H4:} CCE* 
&  a) Caregiver$_{CCE}$ = Caregivee$_{CCE}$ 
&  b) PreStudy$_{CCE}$ < PostStudy$_{CCE}$  
&  c) Caregiving x PrePost|Caregivee$\rightarrow$(+)CCE   \\
 \addlinespace[1ex]
 \textbf{H5:} SE*
&  a) Caregiver$_{SE}$ > Caregivee$_{SE}$ 
&  b) PreStudy$_{SE}$ < PostStudy$_{SE}$ 
&  c) Caregiving x PrePost|Caregivee$\rightarrow$(+)SE   \\
\bottomrule
\multicolumn{4}{l}{ *\textbf{PU}: Power Usage, \textbf{CT}: Community Trust, \textbf{CB}: Community Belonging, \textbf{CCE}: Community Collective Efficacy, \textbf{SE}: Self-Efficacy}  \\
\end{tabular}
\end{table*}

\noindent
\textbf{Survey Constructs and Hypotheses}\\
In the pre-study survey, we measured participants' perceived power usage \cite{sundar2010personalization}, community trust \cite{chouhan_co-designing_2019}, community belonging \cite{kropczynski_examining_2021, carroll_community_2003}, community collective efficacy \cite{carroll_collective_2005, kropczynski_examining_2021}, and self-efficacy \cite{bandura1982self, kropczynski_examining_2021}. All these measures, except power usage, were also later presented in the post-study survey. Our study outcomes were measured on the basis of these constructs. The scale items of these constructs were measured on a 5-point Likert scale from 1 (strongly disagree) to 5 (strongly agree). Below we describe these constructs and their corresponding hypotheses (\autoref{tab:all_hypotheses}).

\textbf{\textit{Power Usage: }}
In the pre-study survey, we measured participants' power usage \cite{sundar2010personalization}, which is defined as their perception of whether they are comfortable with technology in general and are likely to explore all possible features of the technologies they use. In Kropczynski et al.'s work ~\cite{kropczynski_examining_2021}, they demonstrated that the caregivers who support their community in general technology usage had a higher level power usage compared to the caregivees. Therefore, we hypothesize that: 


\textit{\textbf{H1:} Caregivers will report higher levels of power usage than caregivees.}

\textbf{\textit{Community Trust: }} In both the pre- and post-study surveys, we measured our participants' perception of community trust. In this study, we created this construct based on our community oversight model framework that we proposed in \cite{chouhan_co-designing_2019}. We define community trust as an individual's perception of trusting their community to keep their personal information, e.g., apps installed, private and to care for them to provide mobile privacy and security advice and guidance. The scale items are provided in Appendix A. Since our caregivees already received mobile privacy and security support from caregivers of their community before participating in the study, we expect caregivees will have more trust on their communities with their app usage information and will rely on others to receive advice and feedback on their mobile privacy and security. Moreover, as both caregivers and caregivees used the CO-oPS app to help each other with their mobile privacy and security decisions, we expect their community trust to increase at the end of the study. 
Therefore, we hypothesize a main effect of caregiving status prior to using the CO-oPS app, a main effect of the CO-oPS app intervention from pre- to post-study, and an interaction effect between the caregiving status and the CO-oPS app intervention:

\textit{\textbf{H2:}} \textit{a) Prior to using CO-oPS, caregivees will report higher levels of community trust than caregivers,}
\textit{b) Both caregivers and caregivees will report higher levels of community trust after using the CO-oPS app, and}
\textit{c) Caregiving status will moderate the effect of using the CO-oPS app, such that the increase in community trust after using the app will be stronger for caregivees than for caregivers.} 


\textbf{\textit{Community Belonging }} scale \cite{carroll_community_2003} measures an individual's perception of their importance and the degree to which their opinions are valued within their community. In this study, we utilized the pre-validated version of this scale that was used by Kropczynski et al. in \cite{kropczynski_examining_2021}. They showed evidence that community belonging did not differ between tech caregivers and caregivees. This is because both caregivees and caregivers needed higher levels of belonging in their community in order to provide or receive technology support from one another. Therefore, our hypothesis for the main effect of caregiving status before using the CO-oPS app is the null hypothesis. However, as our caregivers and caregivees were connected through the CO-oPS app throughout the study duration and worked together to help one another on their mobile privacy and security decisions, we expect their sense of belonging may increase at the end of the study. Also, since caregivees were supported more by their communities through the CO-oPS app, we expect that they will perceive more belonging after the study than the caregivers. Therefore our hypotheses are:

\textit{\textbf{H3:}} \textit{a) Prior to using CO-oPS, there will be no significant difference in community belonging between caregivers and caregivees,}
\textit{b) Both caregivers and caregivees will report higher levels of community belonging after using the CO-oPS app, and}
\textit{c) Caregiving status will moderate the effect of using the CO-oPS app, such that the increase in community belonging after using the app will be stronger for caregivees than for caregivers.} 



\textbf{\textit{Community Collective Efficacy }} was defined by Carroll et al. \cite{carroll_collective_2005} as the capacity of a group or community for collaboratively performing a shared task. In this study, we used a version of this construct that we validated in previous research \cite{kropczynski_examining_2021}. The community collective efficacy scale assessed how individuals perceived their community's capacity to work together as a team on mobile privacy and security. Prior studies in the space of technology support \cite{kropczynski_examining_2021} showed evidence that caregivers' community collective efficacy was not different than the caregivees. Our study anticipates that the use of the CO-oPS app by caregivers and caregivees to exchange support and guidance on mobile privacy and security will enhance both groups' collective efficacy in managing these issues within their community. Additionally, we expect that our caregivees would receive more help and guidance from the caregivers in making their mobile privacy and security decisions through using the CO-oPS app. Therefore, caregivees' community collective efficacy will become stronger after using the CO-oPS app than the caregivers.

\textit{\textbf{H4:}} \textit{a) Prior to using CO-oPS, there will be no significant difference in community collective efficacy between }
\textit{caregivers and caregivees,}
\textit{b) Both caregivers and caregivees will report higher levels of community collective efficacy after using the CO-oPS app, and}
\textit{c) Caregiving status will moderate the effect of using the CO-oPS app, such that the increase in community collective efficacy after using the app will be stronger for caregivees than for caregivers.} 

\textbf{\textit{Self-Efficacy}}
was originally defined by Bandura \cite{bandura1982self} as an individual's perceived capacity to perform a task. In this study, we employed a pre-validated version of this construct from our prior work \cite{kropczynski_examining_2021}. Their self-efficacy scale measured an individual's perceived ability to manage their own privacy and security. Their work showed evidence that technology caregivers have higher self-efficacy than caregivees. 
In our study, since caregivers provided privacy and security support to their caregivees prior to participating in the study, we expect that our caregivers will also have higher self-efficacy for mobile privacy and security than the caregivees.  Furthermore, since our caregivees used the CO-oPS app to learn from their community members' mobile privacy and security behavior and received feedback and guidance from others on their own mobile privacy practices, we expect that our caregivees' self-efficacy would become stronger after using the CO-oPS app than the caregivers. Therefore, we hypothesized that:

\textit{\textbf{H5:}} \textit{a) Prior to using CO-oPS, caregivers will report higher levels of self-efficacy than caregivees,}
\textit{b) Both caregivers and caregivees will report higher levels of self-efficacy after using the CO-oPS app, and}
\textit{c) Caregiving status will moderate the effect of using the CO-oPS app, such that the increase in self-efficacy after using the app will be stronger for caregivees than for caregivers.}\\


\noindent
\textbf{Data Analysis Approach}\\
In this section, we first describe how we conducted our statistical analyses on the survey data to answer our RQ1 and RQ2 and test the related hypotheses. Then, we describe the quantitative and qualitative analyses approaches in the CO-oPS app usage logs and the interview transcripts to answer our RQ3.

\textbf{\textit{Exploring the Personal Characteristics of Caregivers and Caregivees (RQ1): }}
To investigate the differences in personal characteristics of our caregivers and caregivees, we analyzed their demographic information and the power usage scale that they reported during the pre-study survey. We first categorized our participants as caregivers or caregivees. The individuals whom at least one of their community members reported as the primary person to whom they go to for questions regarding mobile privacy and security, are labeled as caregivers. The rest of the community members whom no one reported as the person they seek mobile privacy and security support from were labeled as caregivees. Next, we investigated whether there were demographic differences between the caregivers and caregivees using chi-square tests. We also measured the difference between our caregivers and caregivees' power usage. We first used Cronbach's alpha \cite{cronbach_coefficient_1951} to verify the internal consistency of this construct, which was 0.85. We then conducted a Shapiro–Wilk test and found that the sum scores of these two groups' power usage were not normally distributed ($ps\!<\!.01$). Therefore, we performed the non-parametric Mann-Whitney U test \cite{noauthor_mannwhitney_2023} between the caregivers' and caregivees' responses to the power usage scale.

\textbf{\textit{Exploring the Difference in Perceived Outcomes between Caregivers and Caregivees (RQ2): }}
To explore differences in perceptual outcomes, e.g., community trust, community belonging, community collective efficacy, and self-efficacy between caregivers and caregivees, we analyzed the pre- and post- study survey responses of our caregivers and caregivees. We first verified the construct validity of these measures using Cronbach's alpha \cite{cronbach_coefficient_1951} and created sum scores to represent each construct. All Cronbach's alphas were greater than 0.85, suggesting a good internal consistency of our measures. We then conducted Shapiro–Wilk tests to identify whether the sum scores of the constructs were normally distributed or not. Since these were not normally distributed, we used the Aligned Rank Transform (ART) tool to perform non-parametric 2 x 2 ANOVAs \cite{Wobbrock2011} and post-hoc contrast (ART-C) tests \cite{Elkin2021} as we anticipated a possible interaction effect between caregiving status and survey completion time. We conducted a 2 x 2 repeated measures ANOVA with mixed design to evaluate how using CO-oPS with the community throughout the study duration affects caregivers' and caregivees' community trust, community belonging, community collective efficacy, and self-efficacy.  \autoref{tab:anovas} presents all these statistical tests for interaction and main effects. The means and standard deviations for all measures are presented in \autoref{tab:means}.

\textbf{\textit{Exploring the Difference in Privacy \& Security Behaviors between Caregivers and Caregivees (RQ3): }}
To examine how our caregivers' mobile privacy and security behaviors were different than the caregivees, we first analyzed our caregivers and caregivees' CO-oPS app usage data. Due to some technical issues with the CO-oPS logging feature, we could not log the first seven communities' app activities and so, the app usage data was collected from only the last fifteen communities (N = 35 caregivers and N=33 caregivees). 
We first prepared our dataset by counting each participant's specific app activities, e.g., count of apps reviewed, messages sent, permissions denied, and apps hidden. We then conducted Shapiro–Wilk tests and found that the sum scores of these app activities were not normally distributed ($ps\!<\!.01$). We then performed the non-parametric Mann-Whitney U test to identify significant differences between the frequencies of the app activities performed by the caregivers and caregivees (\autoref{tab:ttest}). 
We also qualitatively coded the app activity data to examine the types of messages sent (e.g., sought or provided advice, message regarding apps or permissions), types of apps reviewed, apps uninstalled and hidden, and the types of app permissions caregivers and caregivees changed. Lastly, to further explore how differently our caregivers and caregivees perceived these app activities they could perform with the CO-oPS app, we conducted a supplemental qualitative analysis, where we identified themes related to the app activities, but also focused on the differences between our caregivers and caregivees. N=28 caregivers and N=23 caregivees participated in the follow-up interviews. \hyperref[appendix:d]{Appendix D} presents the codes and illustrative quotations for each app activity theme. Participant's quotations are identified by their IDs (e.g., C01P1,..C22P6), caregiving status, age, and gender information. \\

\begin{table*}
  \centering
  \footnotesize
\caption{Community Characteristics}
  \label{tab:groupchar}
  \begin{tabular}{ c c c l l }  \hline
 \textbf{Community ID} & \textbf{Size} & \textbf{Caregiver No.}  & \textbf{Composition} & \textbf{Proximity} \\ \hline
 \multicolumn{5}{c}{} \\[-5pt]  
 C01 & 5 & 3   & partners (2), friends (2), neighbor (1)          & house (2), neighborhood (3) \\ 
 C02 & 5 & 1   & partners (2), parent-teen (2), friend (1)        & house (2), house* (2), town (1)  \\ 
 C03 & 5 & 1   & partners (2), friends (2), co-worker (1)         & house (2), neighborhood (3)   \\ 
 C04 & 2 & 1   & friends (2)                                      & out of town (2)  \\  
 C05 & 5 & 3   & friends (5)                                      & neighborhood(3), town (2)  \\  
 C06 & 6 & 5   & partners (2), friends (2), neighbors (2)         & house (2), neighborhood (3), town (1) \\ 
 C08 & 4 & 2   & partners (2), friend (1), neighbor (1)           & house (2), neighborhood (1), out of town (1) \\ 
 C09 & 5 & 1   & partners (2), friend (1), co-workers (2)         & house (2), town (3)  \\ 
 C10 & 3 & 1   & friend (2), co-worker (1)                        & neighborhood (2), town (1)  \\ 
 C11 & 5 & 3   & friends (5)                                      & neighborhood (4), town (1) \\ 
 C12 & 5 & 4   & friends (3), co-workers (2)                      & house (2), neighborhood (3) \\ 
 C15 & 5 & 1   & partners (2), friend (1), neighbors (2)          & house (2), neighborhood (2), town (1) \\ 
 C16 & 6 & 2   & partners (2), siblings-parent-grand parent (4)   & house (2), neighborhood (2), town (1), out of town (1)\\  
 C17 & 5 & 3   & partners (4), friend (1)                         & house (2), house* (2), town (1)\\ 
 C18 & 5 & 4   & partners (2), teen-parents (3)                   & house (2), house* (3) \\ 
 C19 & 5 & 1   & parent-teen (2), parent-teen-adult child (3)     & house (2), house* (3) \\ 
 C20 & 5 & 3   & friends (5)                                      & neighborhood (3), out of town (2) \\ 
 C21 & 4 & 2   & parents-teens (4)                                & house (4) \\ 
 C22 & 3 & 2   & partners (2), friend (1)                         & house (2), town (1) \\ 
 C23 & 4 & 1   & siblings (2), extended family (2)                & neighborhood (2), out of town (2) \\ 
 C24 & 3 & 2   & adult child-parent-grand parent (3)              & house (3) \\ 
 C26 & 6 & 5   & friends (6)                                      & house (2), neighborhood (3), out of town (1)\\  
 \bottomrule
 \multicolumn{5}{l}{*Another house in the same neighborhood, town, or out of town} \\
\end{tabular}
\end{table*}

\noindent
\textbf{Participant Recruitment and Demographics}\\
We recruited a total of 101 participants that were associated with 22 different groups of caregivers and caregivees. A post hoc power analysis using G*Power \cite{faul_gpower_2007} was used to determine the power of a repeated measures within-between interaction ANOVA F test and found that the power of the test was 99\% for the effect size (Cohen’s d = 0.19, $\alpha$ = 0.05 and total sample size = 101). Participants were recruited through recruitment emails, phone calls, word of mouth, and social media. For each community, we first recruited the initial contact, who completed a screening survey that verified their eligibility based on whether they: 1) reside in the United States, 2) are 13 years or older, 3) have an Android smartphone, 4) are willing to install and use the CO-oPS app, and 5) can participate in this study with at least two other people they knew. 

The screening survey briefly introduced the study and the CO-oPS app. After being screened for eligibility, participants were provided a consent form detailing the study, described as an effort to evaluate the use of the CO-oPS app among their self-formed group and understand how community members can help each other with mobile privacy and security decisions.
\edit{For the teen participants, we required one of their parents or legal guardians to complete the screening survey. If eligible, the parent or guardian then provided consent for their child to participate in the study. The teens themselves received an assent form to agree or deny to participate in our study. On both the consent and assent form, we highlighted the potential benefits of participating in the study as increased awareness of how group members manage app privacy, which might inspire reflection on their personal permission decisions. We also outlined what data will be collected by the CO-oPS app and shared with their group members, and participation could be discontinued at any time. Lastly, we stated that all the data collected would be securely stored in a password-protected Dropbox folder, accessible only to the researchers listed in the IRB.} Participants were then asked to forward the screening survey to their social contacts whom they would like to include as their community members in this study.

\autoref{tab:groupchar} shows community IDs, size, caregiver count, composition (i.e. family, friends, neighbors, coworkers), and proximity of their residences in relation to each other in the community (i.e. same house, neighborhood, town). The size of the communities ranged from 2 to 6, where the majority (68\%, N=15) of the communities had five or more members. Most of the communities (64\%, N=14) had two to five caregivers, and only 36\%, N=8 communities had one caregiver. Most communities (73\%, N=16) were composed of families together with other relationships, e.g., friends, neighbors, co-workers. Among these communities, the most common family dyadic relationships were partners (50\%, N=11), followed by parent-teen (23\%, N=5), parent-adult child (14\%, N=3), siblings (14\%, N=3), and extended families (14\%, N=3). The rest of the communities (N=6, 27\%) consisted mainly of friends. Finally, in terms of proximity types, 
almost all communities (95\%, N=21) had most members living in the same house (73\%, N=16) or same neighborhood (59\%, N=13), but also included members from the same town or out of town.


\begin{table}[hbt!]
\centering
 \footnotesize
\caption{Sociodemographic Characteristics of Caregivers and Caregivees}
  \label{tab:demo}
\begin{tabular}{lcccccccc}
\toprule
                          & \multicolumn{2}{c}{\textbf{Caregivers}} & \multicolumn{2}{c}{\textbf{Caregivees}} & \multicolumn{2}{c}{\textbf{Total}}\\\hline
                        & \textit{n} & \textit{\%} & \textit{n} & \textit{\%} & \textit{n} & \textit{\%} \\ \hline
\textbf{Gender}                  & \textit{}  & \textit{}   & \textit{}  & \textit{}   & \textit{}  & \textit{}   \\ 
 \hspace{3mm}Female  &   23       & 45.1        & 23         & 46        & 46         & 45.5      \\
 \hspace{3mm}Male   & 28         & 54.9        & 27         & 54        & 55         & 54.5        \\
 \hline
\textbf{Age}                     &            &             &            &             &            &             \\

 \hspace{3mm}13-17     & 2          & 4.1        & 4         & 7.7        & 6         & 5.9        \\
 \hspace{3mm}18-24     & 14         & 29.6        & 13         & 25        & 27         & 26.7        \\
 \hspace{3mm}25-34     & 25        & 51       &  24         & 46.2       & 49          & 48.5        \\
 \hspace{3mm}35-44     & 4          & 8.2        & 2          & 3.8         & 6          & 5.9         \\
 \hspace{3mm}45-54     & 4          & 8.2         & 6         & 11.5       & 10         & 9.9         \\
 \hspace{3mm}55-64     & 0          & 0         & 1          & 1.9         & 1          & 1         \\
 \hspace{3mm}65+       & 2          & 3.8         & 0          & 0
 & 2          & 2         \\
 \hline
 \textbf{Ethnicity}               &            &             &            &             &            &             \\ 
 \hspace{3mm}Asian/Pacific Islander          & 36        & 73.5         & 36          & 69.2         & 72          & 71.3         \\
 \hspace{3mm}Black/African American             & 6        & 12.2        & 7       & 13.5        & 13         & 12.8        \\
 \hspace{3mm}Hispanic/Latino    & 4         & 7.8        & 4         & 8        & 8         & 7.9        \\
 \hspace{3mm}White/Caucasian    & 5         & 9.8        & 3         & 6       & 8         & 7.9        \\
 \hline
\textbf{Education}               &            &             &            &             &            &             \\ 
 \hspace{3mm}Primary School          & 3        & 5.9         & 5          & 10         & 8          & 7.9         \\
 \hspace{3mm}High School             & 2        & 4.1        & 3        & 5.8        & 5         & 5.0        \\
 \hspace{3mm}College (Associate)     & 3         & 6.1        & 3         & 5.8        & 6         & 5.9        \\
 \hspace{3mm}College (Bachelor)     & 25         & 51        & 15         & 28.8       & 40         & 39.6        \\
 \hspace{3mm}Masters                 & 16       & 31.3        & 20         & 40        & 36         & 35.6        \\
 \hspace{3mm}Doctoral/Professional   & 1         & 2         & 5          & 9.6        & 6          & 5.9         \\

\bottomrule
\end{tabular}
\end{table}

\section{Results}
\noindent
\textbf{Personal Characteristics of Caregivers and Caregivees}\\ 
We characterize our caregivers and caregivees based on their demographic characteristics and level of power usage to identify the ways in which these two groups differ. Out of the 101 participants, N=51 (50\%) were classified as caregivers, and N=50 (50\%) as caregivees. \autoref{tab:demo} illustrates the gender, age groups, ethnicity, and education of our caregivers and caregivees. Overall, we found no significant differences between the caregivers and caregivees based on any of these demographic characteristics. 
Next, we examine power usage and find that the caregivers (mean=3.87, SD= 0.50) reported significantly higher levels of power usage than the caregivees' (mean = 3.61, SD = 0.51), with statistical significance (\textit{p} = 0.04). This suggests that our caregivers had more expertise in technology, compared to the caregivees. Therefore, our hypothesis H1 is supported. \\

\begin{table}[t]
 \footnotesize
 \centering
  \caption{Summary of the Statistical 2 x 2 ANOVAs}
  \label{tab:anovas}
\begin{tabular}{lllcc}  
\hline
 \textbf{Effect Location} & \textit{F (1, 99)} & \textit{p-val} & $\eta^2_p$ & \\ \hline

\multicolumn{4}{l}{\textbf{\underline{Community Trust:}}} &\\
 
\hspace{9mm}Caregiving  & 38.21*** & \textbf{<0.001} & 0.28 \\
\hspace{9mm}Pre-Post  & 13.77*** & \textbf{<0.001} & 0.12 \\
 \hspace{9mm}Interaction & 3.51 & 0.06 & 0.034 \\   \addlinespace[1ex]

\multicolumn{4}{l}{\textbf{\underline{Community Belonging:}}} &\\
  
\hspace{9mm}Caregiving  & 26.26*** & \textbf{<0.001} & 0.21 \\
 \hspace{9mm}Pre-Post & 2.85 & 0.10 & 0.03 \\
\hspace{9mm}Interaction & 0.03 & 0.86 & 0 \\   \addlinespace[1ex]

\multicolumn{4}{l}{\textbf{\underline{Community Collective Efficacy:}}} & \\ 
\hspace{9mm}Caregiving & 92.55*** & \textbf{<0.001} & 0.48 \\
 \hspace{9mm}Pre-Post  & 81.39*** & \textbf{<0.001} & 0.45 \\
 \hspace{9mm}Interaction & 18.87*** & \textbf{<0.001} & 0.16 \\   \addlinespace[1ex]

\multicolumn{4}{l}{\textbf{\underline{Self-Efficacy:}}} & \\ 
\hspace{9mm}Caregiving  & 34.73*** & \textbf{<0.001} & 0.26 \\
 \hspace{9mm}Pre-Post & 108.71*** & \textbf{<0.001} & 0.52 \\
 \hspace{9mm}Interaction & 30.19*** & \textbf{<0.001} & 0.23 \\ 

\bottomrule
\multicolumn{4}{l}{*p\textless{}.05; **p\textless{}.01; ***p \textless .001} & \\

\end{tabular}
\end{table}

\begin{table}[h!]
 \footnotesize
\caption{Means and Standard Deviations of All Constructs.}
\begin{tabular}{llcccc}
\toprule
   &  & \multicolumn{2}{l}{\textbf{Caregiver}} & \multicolumn{2}{l}{\textbf{Caregivee}}    \\ \cline{3-6} 
& \textbf{} & \textbf{M1} & \textbf{SD1} & \textbf{M2} & \textbf{SD2}  \\ \hline
\textbf{\textit{Pre-Study}} & Community Trust & 4.40 & 0.55 & 3.60 & 0.88 \smallskip \\
& Community Belonging & 4.30 & 0.49 & 3.86 & 0.66 \\
& Community Collective Efficacy & 4.36 & 0.40 & 3.30 & 0.68\\
& Self-Efficacy & 4.33 & 0.36 & 3.59 & 0.64  \\

\textbf{\textit{Post-Study}} & Community Trust & 4.54 & 0.58 & 4.02 & 0.72 \\ 
& Community Belonging & 4.44 & 0.61 & 3.98 & 0.67 \\
& Community Collective Efficacy & 4.55 & 0.41 & 4.04 & 0.59\\
& Self-Efficacy & 4.58 & 0.36 & 4.40 & 0.55  \\
\bottomrule
\end{tabular}
\label{tab:means}
\end{table}

\begin{figure}[t]
\centering
\begin{subfigure}[t]{.49\textwidth}\centering
  \includegraphics[width=\columnwidth]{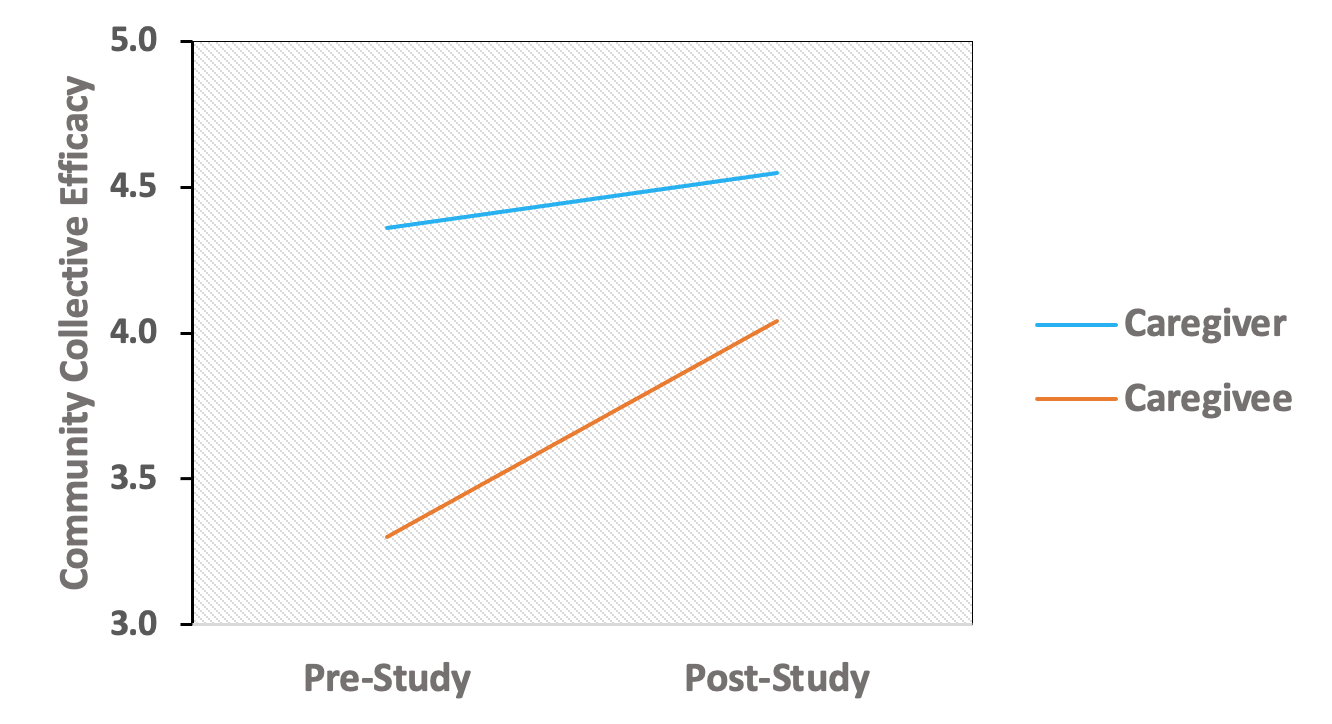}
  \caption{}
\end{subfigure}%
\\
\begin{subfigure}[t]{.49\textwidth}\centering
  \includegraphics[width=\columnwidth]{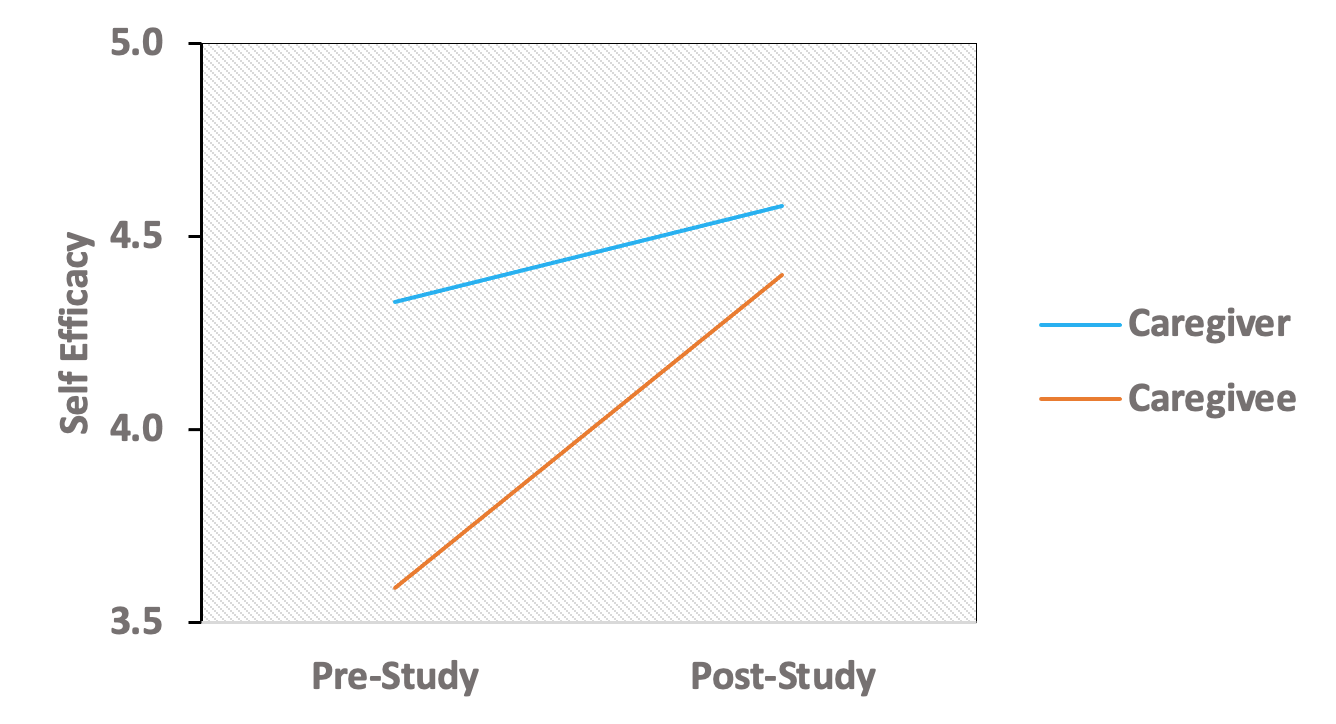}
  \caption{}
\end{subfigure}
\caption{Relationship between the CO-oPS app usage across pre-post study and (a) Community Collective Efficacy (b) Self-Efiicacy for caregivers and caregivees }~\label{fig:interactions}
\end{figure}

\noindent
\textbf{Perceived Outcomes of Caregivers and Caregivees}\\
\autoref{tab:anovas} shows the statistical tests for interaction and main effects of caregiving status and pre-post study. The means and standard deviations of the caregivers and caregivees' pre- and post-study measures are provided in \autoref{tab:means}. 

\textbf{\textit{Community Trust: }}For community trust, we found a significant main effect for both caregiving status (F(1, 99) = 38.21, \textit{p} < 0.001) and pre-post study, (F(1, 99) = 13.77, \textit{p} < 0.001), with no statistically significant interaction (F(1, 99) = 3.51, \textit{p} < 0.06). This suggests that community trust was higher for caregivers than caregivees prior to using the CO-oPS app (opposite of our hypothesis H2a). Both the caregivers and caregivees reported higher levels of community trust after using the CO-oPS app, supporting our hypothesis H2b. Lastly, the caregiving effect on community trust was not greater in the post-study than in the pre-study, indicating that the positive main effect of using the CO-oPS app on caregivees' community trust was not stronger than caregivers'. Therefore, our hypothesis H2c was not supported.

\textbf{\textit{Community Belonging: }}
The pre-study community belonging was higher for caregivers than caregivees, with F(1,99) = 26.26, \textit{p} < .001, suggesting that our null hypothesis H3a was not supported. However, the main effect of the pre-post study and the interaction effect were nonsignificant, suggesting that the increase in caregivees and caregivers' community belonging from pre-to-post study was not significant. Therefore, both hypotheses H3b and H3c were not supported. 

\textbf{\textit{Community Collective Efficacy: }}
For community collective efficacy, there was a significant main effect for caregiving status, F(1, 99) = 92.55, \textit{p} < .001, indicating that the pre-study community collective efficacy was higher for caregivers compared to caregivees (H4a was not supported). The main effect of pre-post study for community collective efficacy also yielded an F ratio of F(1, 99) = 81.4, \textit{p} < .001, with a significant interaction, F(1,99) = 18.87, \textit{p} < .001, suggesting that it increased for both caregivers and caregivees from pre to post study, as illustrated in \autoref{fig:interactions}a, and the positive main effect of the use of the CO-oPS app (pre-post study) on community collective efficacy was also stronger for caregivees than for caregivers. So, hypotheses H4b and H4c were supported.

\textbf{\textit{Self-Efficacy: }}
In terms of self-efficacy, we found a significant main effect for both caregiving status, F(1, 99) = 34.73, \textit{p} < 0.001, and pre-post study, F(1,99) = 108.71, \textit{p} < 0.001, with a statistically significant interaction, F(1,99) = 30.19, \textit{p} < 0.001. This suggests that self-efficacy was higher for caregivers than caregivees before the study and increased for both caregivers and caregivers from pre-study to post-study test, as shown in \autoref{fig:interactions}b. Also, the positive main effect of using the CO-oPS app (pre-post study) on self-efficacy was stronger for caregivees than for caregivers. This confirms that all three of our hypotheses H5a, H5b, and H5c are supported. \autoref{tab:hypotheses} summarizes the hypotheses tested from our statistical model.\\

\begin{table*}[t]
 \footnotesize
 \centering
  \caption{Hypotheses Testing Results}
  \label{tab:hypotheses}
\begin{tabular}{llllc}  
\hline
 \textbf{ Constructs}  & \textbf{ Main Effect of Caregiving ($\boldsymbol{\checkmark/X})^+$} &  \textbf{Main Effect of CO-oPS $(\boldsymbol{\checkmark/X})^+$} &  \textbf{Interaction Effect ($\boldsymbol{\checkmark/X})^+$} &  \textbf{ Supported}  \\ \hline

\addlinespace[1ex]

 \textbf{H1:}$PU^*$   
&  Caregiver$_{PU}$>Caregivee$_{PU} \boldsymbol{\checkmark}$
&  -- 
&  --   
& Yes\\
 \addlinespace[1ex]
 
 \textbf{H2:} $CT^*$ 
&  a) Caregiver$_{CT}$<Caregivee$_{CT} \textbf{X}$
&  b) PreStudy$_{CT}$<PostStudy$_{CT} \boldsymbol{\checkmark}$
&  c) Caregiving x PrePost|Caregivee$\rightarrow$(+)CT \textbf{X}
& Partially \\
 \addlinespace[1ex]

 \textbf{H3:} $CB^*$ 
&  a) Caregiver$_{CB}$=Caregivee$_{CB} \textbf{X}$
&  b) PreStudy$_{CB}$<PostStudy$_{CB} \textbf{X}$
&  c) Caregiving x PrePost|Caregivee$\rightarrow$(+)CB \textbf{X} 
& No\\
 \addlinespace[1ex]

 \textbf{H4:} $CCE^*$ 
&  a) Caregiver$_{CCE}$=Caregivee$_{CCE} \textbf{X}$ 
&  b) PreStudy$_{CCE}$<PostStudy$_{CCE} \boldsymbol{\checkmark}$
&  c) Caregiving x PrePost|Caregivee$\rightarrow$(+)CCE $\boldsymbol{\checkmark}$ 
& Partially \\
 \addlinespace[1ex]

 \textbf{H5:} $SE^*$
&  a) Caregiver$_{SE}$>Caregivee$_{SE} \boldsymbol{\checkmark}$
&  b) PreStudy$_{SE}$<PostStudy$_{SE} \boldsymbol{\checkmark}$ 
&  c) Caregiving x PrePost|Caregivee$\rightarrow$(+)SE $\boldsymbol{\checkmark}$   
& Yes\\

\bottomrule

 \multicolumn{5}{l}{ $\boldsymbol{^+\checkmark/X}$: Hypothesis was supported or not; $^*$\textbf{PU}: Power Usage, \textbf{CT}: Community Trust, \textbf{CB}: Community Belonging, \textbf{CCE}: Community Collective Efficacy, \textbf{SE}: Self-Efficacy} \\
\end{tabular}
\end{table*}

\begin{table}[t]
 \footnotesize
 \centering
  \caption{Mann-Whitney U Tests of App Activities}
  \label{tab:ttest}

\begin{tabular}{lccccclc}  
\hline
& \multicolumn{2}{l}{\textbf{Caregiver}} & \multicolumn{2}{l}{\textbf{Caregivee}} &  &   \\ \cline{2-5} 
& \textit{M1} & \textit{SD1} & \textit{M2} & \textit{SD2} & \textit{W-val} & \textit{p-val} \\ \hline
\textbf{Apps reviewed } & 10 & 16.71 & 7  & 27.43 & 244 &  0.689 \\
\textbf{Messages sent} & 1 & 0.80 & 1  & 0.97 & 320.5 &  0.952 \\
\textbf{Apps uninstalled} & 1 & 4.33 & 3  & 11.25 & 264 &  0.321 \\
\textbf{Permissions denied} & 3 & 3.00 & 3  & 2.45 & 67.5 &  0.263 \\
\textbf{Apps hidden } & 3 & 13.18 & 11  & 21.25 & 278.5* &  \textbf{0.043} \\
\bottomrule
*p\textless{}.05; **p\textless{}.01; ***p \textless .001 & & & & & &  \\
\end{tabular}

\end{table}

\noindent
\textbf{Privacy \& Security Behaviors of Caregivers and Caregivees}\\
As described in our Methods section, we conducted both quantitative and qualitative analysis on participants' app usage logs and interview transcripts for our RQ3. 
\edit{On average, caregivers spent 38 minutes on the CO-oPS app over the duration of the study, with individual usage ranging from 21 minutes to 1 hour and 31 minutes. In comparison, caregivees spent an average of 29 minutes using the app during the study, with individual usage ranging from 19 minutes to 58 minutes.} Below we present the results for how the caregivees and caregivers are different based on the app activities they performed on the CO-oPS app and also, how they perceived these app activities differently to support them in providing or receiving mobile privacy and security care from their communities.

\textbf{\textit{Apps Reviewed: }}
As shown in \autoref{tab:ttest}, while the median number of apps reviewed by caregivers was slightly higher, we found no significant difference (\textit{p} = 0.689) between caregivers (median = 10, SD= 16.71) and caregivees (median = 7, SD= 27.43) based on the number of the apps they reviewed. When we further investigated who reviewed whose apps more (\autoref{tab:appsreviewed}), we found that caregivees' apps were reviewed more compared to the caregivers'. For example, caregivers reviewed a total count of 141 apps of other caregivers, with an average of 5.6 apps, while reviewing almost twice as many apps from caregivees (total = 263, avg = 10.5). Similarly, caregivees reviewed 278 apps from other caregivees (avg = 13.2), almost twice as much as caregivers' apps reviewed (total = 144, avg = 6.9). 

\begin{table}[t]
 \footnotesize
\centering
 \caption{Apps Reviewed}
  \label{tab:appsreviewed}

\begin{tabular}{lcccccc}
\hline
& \multicolumn{3}{c}{\textbf{Caregivers's apps}} & \multicolumn{3}{c}{\textbf{Caregiveees' apps}} \\\cline{2-7}
& \textit{n} & \textit{avg.} & \textit{\%} & \textit{n} & \textit{avg.} & \textit{\%} \\ \hline 
 \textbf{Caregivers}  & 141     & 5.6 & 49.5    & 263  & 10.5 &   48.6   \\
 \textbf{Caregivees}  & 144 &  6.9  &  50.5    & 278  & 13.2 & 51.4   \\
 \bottomrule
\end{tabular}
 
\end{table}

From our data analysis of the interview, we found that more caregivees felt that using the CO-oPS app with their groups helped them receive valuable feedback. A majority of caregivees 
mentioned that having other people review their apps helped them \textbf{identify risky app permissions}, which was made possible through the feedback they received from others.  For example, C02P2, a 51-year-old male caregivee, said: 
\textit{“It made them [community] see how I manage my apps and their data access. Like, I didn't give that much thought to my permissions, but they could see what I did and just tell me what I need to fix. This is great because you can be aware of the fishy permissions or the apps you have.”} Most caregivees also felt that this feature gave them \textbf{peace of mind} that there was someone else in the community who would watch over their privacy and security. Caregivees frequently mentioned specific caregivers in their community on whom they would rely for mobile privacy and security advice. Having their caregivers included in the CO-oPS app who could keep an eye on their apps and permissions made caregivees less concerned. 
Interestingly, caregivers felt that this feature that allowed every community member to review everyone else's apps and permissions helped \textbf{distribute community responsibility}. This is because people often use numerous apps on their phones, and reviewing each of their permissions can be tedious for one person. Here, a few caregivers also mentioned that not all people in a community have the same level of knowledge of recognizing the dangerous permissions, and therefore having multiple people review someone's apps would be beneficial. For instance, C19P6, 52-year-old female caregiver, said: 
\textit{“
It's nice, because someone can have multiple feedback and use their judgement whether they keep the permission or not. Also, because some people in the community can not give good feedback, but others can. So, its good to have all people go through and inform.”}


\begin{table*}[tbh]
 \footnotesize
\centering
 \caption{Types of Messages Sent}
  \label{tab:messagesent}
\begin{tabular}{lcccccc|cccccc|cccccc}
\toprule
& \multicolumn{3}{l}{\textbf{to Caregivers}} & \multicolumn{3}{l|}{\textbf{to Caregiveees}} & \multicolumn{3}{l}{\textbf{about apps}} & \multicolumn{3}{l|}{\textbf{about permissions}}  & \multicolumn{3}{l}{\textbf{provided advice}} & \multicolumn{3}{l}{\textbf{sought advice}} \\\cline{2-19}
& \textit{n} & \textit{avg.} & \textit{\%} & \textit{n} & \textit{avg.}  & \textit{\%} & \textit{n} & \textit{avg.} & \textit{\%} & \textit{n} & \textit{avg.}  & \textit{\%} & \textit{n} & \textit{avg.} & \textit{\%} & \textit{n} & \textit{avg.}  & \textit{\%} \\ \hline 
 \textbf{Caregivers}  & 17 & 1.4 & 54.8       & 21 &  1.9 &   42.9
             & 11 & 0.3 & 45.8       & 27 &  1.1 &   48.2   
             & 28 & 1.2 & 45.9       & 10 &  0.3 &   52.6  \\
 \textbf{Caregivees}  & 14 & 0.6 & 45.2       & 28 &  1.9 &   57.1 
             & 13 & 0.4 & 54.2       & 29 & 1.3  &   51.8     
             & 33 & 1.4 & 54.1       & 9  & 0.3  &   47.4  \\
 \bottomrule
\end{tabular}
\end{table*}

\textbf{\textit{Messages Sent: }}
As shown in \autoref{tab:ttest}, we found no significant difference (\textit{p} = 0.952) in the total count of messages sent by caregivers (median = 1, SD = 0.8) and caregivees (median = 1, SD = 0.97). However, both caregivers and caregivees sent slightly more messages to caregivees than to caregivers. 

Qualitative analysis on the types of messages sent also gave us important insights (\autoref{tab:messagesent}). For example, the majority of the messages sent by both caregivers and caregivees were about the permissions granted. 
For instance C17P3, 32-year-old male caregiver, sent this message to C17P4, 32-year-old female caregivee: \textit{"your Bofa camera permission is granted which you should denied access."}. Both caregivers and caregivees sent more messages to provide advice, compared to the count of messages sent to seek advice, and there were no differences between caregivers and caregivees' for this. The qualitative analysis of our interview transcripts gave us further insights into how the caregivers and caregivees overall felt about the CO-oPS app's messaging feature that facilitated exchanging advice and guidance on their mobile privacy and security. Most caregivers mentioned that \textbf{initiating discussions} about the installed apps of others and granted permissions became easier with the CO-oPS app, even when community members are not close. They often mentioned that they would usually hesitate to start conversations about the apps installed on someone else's devices. But since the CO-oPS app allowed them to review others' apps and then communicate their opinion through the messaging feature, it addressed that hesitation. For example, a 28-year-old male caregiver, C06P6, said:
\textit{“Look, this is really an awkward topic, especially when your community is not your family. You can not just tell someone out of nowhere, hey your x app is accessing microphone. This app has so many features to communicate this awkward topic. So, telling someone about their personal apps and their settings is a bit less difficult with it."} Caregivees, on the other hand, felt that the CO-oPS app made it easy to \textbf{seek advice and guidance}. Interestingly, most of these caregivees said that they either used other messaging tools, e.g., Whatsapp, Facebook Messenger, text message, etc., or just talked in person to ask questions because they utilized those tools for their daily communication or because they lived in same neighborhood or house. However, most of our caregivers still saw value in the in-app messaging feature, as it helped \textbf{keep the conversation specific} to the topic of mobile privacy and security. They often said that using other messaging tools would not help much as the advice given would be lost among their other general conversations. Therefore, they found the messaging feature useful in preserving the advice given on the apps or permissions. 


\begin{table}[t]
 \footnotesize
\centering
 \caption{Apps Uninstalled and Permissions Denied}
  \label{tab:permissionsdenied}
\begin{tabular}{lcccccc}
\toprule
& \multicolumn{3}{l}{\textbf{uninstalled apps}} & \multicolumn{3}{l}{\textbf{changed permissions}} \\\cline{2-7}
& \textit{n} & \textit{avg.} & \textit{\%} & \textit{n} & \textit{avg.} & \textit{\%} \\ \hline 
 \textbf{Caregivers}             & 12  & 0.3 &  29.3       & 38 & 2.3 & 43.7  \\
\textbf{Caregivees}              & 29  & 2.8 &  70.7       & 49 & 2.6 & 56.3  \\
 \bottomrule
\end{tabular}
\end{table}

\textbf{\textit{Apps/Permissions Changed: }}
We now describe how our caregivers and caregivees' mobile privacy and security behaviors changed at the end of the study. We specifically looked into caregivers and caregivees' permission changes through the CO-oPS app and the changes in their installed app list (\autoref{tab:permissionsdenied}). As illustrated in \autoref{tab:ttest}, we found no significant differences between the caregivers and caregivees based on both the apps uninstalled (\textit{p}=0.321) and permissions denied (\textit{p}= 0.263). Qualitative analysis of the app log data revealed that caregivees mostly uninstalled gaming apps, workout, and dictionary apps, whereas the caregivers uninstalled finance, mobile payment, and shopping apps. Our caregivers mainly denied contacts and microphone permissions, while caregivees mainly denied storage and camera. However, both caregivers and caregivees showed a similar trend in denying location permissions. 

During interviews, all caregivees said that \textbf{comparing} their own app permissions with others inspired them to take the initiative and change the permissions. To this end, they often mentioned the CO-oPS feature that allowed users to view the count of community members who allowed or denied the permissions of an app, which helped them learn whether the permissions are necessary for a specific app and then change the permission accordingly. For instance, C12P4, a 25-year-old male caregivee, said: 
\textit{“I can see what kind of permissions they're getting to their apps and how many. I saw that some of my friends didn't get permission to Facebook, on their location or something. So I see that in my app, I already give the permission of location. So after seeing that, my other friends haven't given that permission, I removed my location permission from my Facebook. ”} Caregivees also often felt that they \textbf{learned how to change} their app permissions. Here, caregivees often mentioned the CO-oPS app feature that allowed users to easily navigate to the Android settings to turn off/on the app permissions, which enabled them to change the permissions easily.  Our caregivees also often said that learning about \textbf{others' permission changes encouraged} them to make the changes for their own. To this end, they mostly brought up the CO-oPS' community feed where they would get posts about the app permission changes made by their community members. 



\textit{\textbf{Apps Hidden:}}
The CO-oPS app allows users to hide apps from their community members that they do not wish others to know about. Interestingly, we found that our caregivees hid more apps (Median = 11, SD = 21.25) than the caregivers (median = 3, SD = 13.18), and the difference was statistically significant (\textit{p} = 0.043). 
In total, 21 caregivees hid \textit{N} = 192 apps (68.8\%), which was approximately double the number hidden by 13 caregivers (31.2\%, \textit{M} = 87). \textcolor{black}{Upon further analysis of the types of apps hidden, we discovered that caregivees predominantly hid social media apps (38\%, \textit{n} = 73), including messaging and video apps such as Instagram, WhatsApp, Youtube, and TikTok, as well as online shopping apps (21\%, \textit{n} = 41) such as Amazon, Walmart, Macy's, Target, and Kohl's. In contrast, caregivers mostly hid gaming apps (32\%, \textit{m} = 28) like Pokemon Go, Wordscapes, Clash of Clans, and Candy Crush, and video streaming apps (25\%, \textit{m} = 22) such as Disney+, Netflix, Peacock, and HBO Max. Additionally, both caregivers (15\%, \textit{m} = 13) and caregivees (11\%, \textit{n} = 22) hid financial apps including CashApp, Venmo, Zelle, and various banking apps.}   

Our qualitative analysis of the interview data gave us insights to further unpack why caregivees hid their apps more than caregivers. In general, caregivees were more \textbf{concerned about their privacy} on their app usage than the caregivers. All caregivees mentioned that the CO-oPS app allowed their community members to review their personal apps that they did not feel comfortable about. Caregivees often mentioned that while they generally liked the idea that others can review their apps and provide feedback, they would still feel uncomfortable sharing their apps with some of their community members. They also often felt uncomfortable when reviewing others' mobile apps installed, as it was perceived as violating others' privacy. For instance, C09P2, a 23-year-old male caregivee, said:
\textit{“The nature of this app is showing what apps we have installed on all devices to everyone else. The nature of the app, also kind of decreases privacy for your own and for others. In sum, individually. But I guess the idea is that you have a group of people you have to have trust somewhat so you can get help. But still it kind of gives a discomfort.”} Caregivees also often specifically mentioned that when the community consisted of \textbf{mixed relationships}, their concerns about privacy could be aggravated. For example, they would not want their friends to monitor their family members' apps or vice versa. For instance C15P2, a 33-year-old female caregivee, said: 
\textit{“You don't want your friends to be in the same community with your parents. Because you don't want your parents to see some of your apps or your friends to see some of your other apps, right? So yeah, I guess that was my main concern. … I also feel like I don't think my friends should be seeing my parents' apps, either.”} Caregivers, on the other hand, expressed concern about this app hiding feature because when caregivees hide their apps it would \textbf{defeat the primary purpose} of the CO-oPS app. Caregivers often brought up the importance of the community members' personal privacy on their app usage, but they also felt that providing feedback would be challenging if a community member hid most of their apps. To this end, caregivers frequently gave examples of vulnerable members of their families who do not have much expertise of mobile privacy and security, e.g., children, parents, grand parents, etc. 


\section{Discussion}
In this study, we examined whether and how collaborative mechanisms for mobile privacy and security support caregiving relationships, and how caregiving status influences the outcomes of using the CO-oPS app. Overall, we found that caregivers and caregivees did indeed have different outcomes, exhibited different behaviors, and ultimately experienced the app in unique ways. In the following, we describe the implications of our findings in relation to previous work and provide recommendations for implementing community-based solutions that support community-based caregiving. \\

\noindent
\textbf{Power Use as Difference between Caregivers and Caregivees}\\
Interestingly, we did not find significant differences between our caregivers and caregivees based on their demographic characteristics. Although this may partially be due to low sample-sizes across some groups (e.g., teens and older adults), anecdotally, we also found that older adults and youth in our study acted as caregivers. These findings run counter to common narratives in research  (c.f., ~\cite{franz_gender_2019,  czaja_older_2008, kropczynski_examining_2021, tanni2024lgbtq, akter_it_2023,  kiesler_troubles_2000, correa_brokering_2015}), where older adults tend to seek support and younger generations provide the support to their families. Therefore, we give networked privacy and security researchers a word-of-caution -- not to unintentionally stereotype caregivers/caregivees based on age or other demographics.  

On the other hand, we hypothesized and confirmed that caregiving status differed significantly by power usage (i.e., higher for caregivers). This finding is important in that it serves to validate that individuals' self-reported perceptions of who provides oversight aligned well with the level of power usage of these individuals, suggesting that caregiving for mobile privacy and security is related to the competence and desire to use technology to its fullest \cite{moses_examining_2021}. As we have previously noted in \cite{akter_evaluating_2023}, CO-oPS users often referred to tech savvy members of their community because they were more knowledgeable about mobile privacy and security. Therefore, future research could utilize power usage as a proxy measure to differentiate between caregivers and caregivees for mobile privacy and security within community-based settings -- or it could simply ask individuals from whom they receive oversight as we did.\\



\noindent
\textbf{Amplified Effects of Community Oversight for Caregivees}\\
A key finding from our study was that the community-based approach of using CO-oPS to provide joint oversight was beneficial to \textit{both parties}, but it was even \textit{more beneficial} for caregivees. We saw a significant moderating effect of caregiving status, such that self-efficacy and community collective efficacy increased more for caregivees than caregivers from pre- to post-study. \edit{Interestingly, while caregivees received oversight from their community members, as our RQ3 results indicated, they also provided oversight to others, which may have contributed to their enhanced self-efficacy.} Overall, our findings confirmed our hypothesis: people receiving care knew less about privacy and security than those providing care, likely because they had more to learn and needed more guidance. Consequently, our study provides strong empirical evidence for how community-based approaches can disproportionately benefit those in need of help \cite{wan_appmod_2020, anaraky2019testing, akter_it_2023, kropczynski_towards_2021} in ways that benefit the community as a whole. Therefore, we urge HCI researchers to continue to emphasize the importance of collaboration and community across all aspects of our scholarship. However, our data revealed several trends that ran contrary to the effects we hypothesized, which also offer valuable insights. In particular, caregivers reported significantly higher levels of community trust and belonging than caregivees, when we expected higher levels of trust from caregivees due to their reliance on caregivers. These unanticipated findings may partially be explained by our RQ3 results, which revealed that caregivees had more privacy concerns; and therefore, hid more apps than caregivers. This surfaces a potential tension between caregivees and caregivers due to their differing roles. 

As such, a critically important takeaway is to make sure these technologies for community-based oversight do not unintentionally create oppressive power hierarchies that could harm vulnerable populations. HCI researchers have highlighted the potential pitfalls of surveillance-based technologies among marginalized individuals (e.g., children, older adults, victims of intimate partner violence (c.f., ~\cite{leitao_anticipating_2019, parkin_usability_2020, rodriguez-rodriguez_towards_2020, slupska_threat_2021, mcdonald_privacy_2022, noauthor_ipv_2022}). Therefore, when attempting to protect caregivees from digital privacy and security threats, it is important to assess whether protection is needed from the caregivers themselves. For example, while allowing users to hide apps within CO-oPS provides a safety mechanism against unwanted oversight, caregivees could still be unduly punished for such behavior outside of the app by caregivers who insist on wielding coercive control. One approach for mitigating users' personal privacy, safety and security concerns is through the careful design of these community-based collaborative mechanisms. Future research needs to find effective ways to increase caregivees' community belonging and trust to intrinsically motivate them to use the app voluntarily, rather than making community participation compulsory. \\

\noindent
\textbf{Caregivers and Caregivees Behave Differently}\\
In terms of the behavioral differences between our caregivers and caregivees, we only found statistically significant differences in the number of apps they hid from their community members. Perhaps, this is likely because we asked both our caregivers and caregivees to perform the same weekly tasks. However, qualitatively, we did see some differences in their privacy behaviors. Although our interview results showed a distinction between caregivers and caregivees in terms of initiating discussions to provide guidance and support and asking questions or seeking advice, our log results showed that both groups provided advice to the other caregivers and caregivees of their communities. This is because the collaborative community-based mechanisms of the CO-oPS app were designed to provide an equal footing for everyone in the community, regardless of their roles and expertise. Yet, our caregivers tended to review more apps, while caregivees changed relatively more app permissions based on the oversight received. In retrospect, caregivees providing advice may cause more harm than benefit, as they may lack the necessary skills to provide mobile privacy and security advice to others. This may also potentially cause tensions within the community, as caregivers might ignore their advice \cite{akter_it_2023} and the other caregivees may receive potentially wrong advice \cite{akter_from_2022}. 

Therefore, a key implication of our findings is that future research might find it advantageous to discriminate between study participants in terms of their assigned tasks, so that they are more appropriate for their individual roles. 
This may resolve the tensions and discomfort that we observed among many of our caregivees when asking them to play the role of a caregiver. Such consideration would likely improve participants' experience by accounting for differing roles earlier in the design process, rather than post hoc, as was a limitation of our study. Importantly, these findings suggest that the community-based approaches should be designed to support different caregiving roles in heterogeneous communities. \\




\noindent
\textbf{Implications for Design}\\
Our results demonstrate how caregivers and caregivees are uniquely different in communities and how their perceptions differed when they used a community-based mobile privacy and security mechanism to exchange caregiving. 
\edit{
The insights gained from the use of the CO-oPS app offer real-world applications not only for similar tools but also for the design of broader strategies aimed at helping users safeguard their privacy across various technological domains. These insights extend beyond mobile apps to other technology environments where users share personal information with third parties, such as smart home devices \cite{alghamdi_codesigning_2023, alghamdi_misu_2022}, social media platforms \cite{bhagavatula_adulthood_2022, alghamdi2022webprototype, ulusoy_panola_2021} and websites \cite{mcdonald_citizens_2021}. For example, collaborative community-based mechanisms can be adapted to assist individuals in making informed decisions about the data collection by IoT devices, as demonstrated in the work of Emami-Naeini et al. \cite{emami_influence_2018}. Similarly, McDonald et al. \cite{mcdonald_building_2021} explored the design of technologies that enable collaboration among loved ones in managing cybersecurity, reflecting the broader applicability of community-driven approaches in enhancing privacy and security practices across different technological landscapes.}
Below we discuss the features and mechanisms that need to be in a collaborative oversight tool to support these differences of roles and expertise in communities. 

\textbf{\textit{Identifying Caregivers in Communities: }}
Our study examined the differing roles of caregivers and caregivees in community oversight, albeit in a post hoc fashion. To gain deeper insights, future researchers and designers should consider integrating on-boarding features aimed at distinguishing caregivers from individuals primarily in need of care. This could be achieved by prompting community members to self-identify their caregiving roles during the sign-up process. Another approach could involve identifying power users within the community. However, the power use scale utilized in our study comprised 22 items \cite{sundar2010personalization}, which may pose feasibility challenges for users. Therefore, we recommend utilizing a condensed version of this scale, such as the one developed by Sundar et al. in \cite{sundar_measure_2016}, which consists of 12 items.

\textbf{\textit{Addressing when Caregivers are Absent: }} 
In some cases, communities might lack technological expertise needed to provide care. To mitigate this problem, researchers and designers may need to explore ways to infuse external expertise in the community, as we have previously suggested in \cite{chouhan_co-designing_2019}. Another possibility is that rather than bringing external experts into a community, it might be advantageous to educate community members. In real-world settings, some caregivees may eventually learn to take the lead to provide oversight within their communities. Identifying these transitions over time is also important. Therefore, we recommend incorporating smart caregiving detection mechanisms into the application to recognize caregiving behaviors, such as sending messages containing privacy and security advice. This may require the integration of real-time messaging APIs to discern the intent behind messages, such as whether they are intended to seek or offer advice. 
\textbf{\textit{Designing for Different Caregiving Roles:}}
We recommend that app designers adopt a nuanced approach when developing community oversight tools to accommodate the distinct roles of caregivers and caregivees, rather than unequivocally treating them as interchangeable entities. Utilizing methodologies such as storyboarding and creating user personas can help to fully understand and address the unique needs and expectations associated with these roles. However, it is imperative to recognize and mitigate potential tensions arising from power dynamics and disparities in privacy and security knowledge. To effectively navigate these complexities, we suggest incorporating methodologies such as value-sensitive design \cite{friedman_value_2002} or participatory design ~\cite{spinuzzi_methodology_2005, muller_participatory_1993}, which prioritize the integration of diverse perspectives and values throughout the design process. These methodologies would help facilitate meaningful engagement of all stakeholders, ensuring that the resulting community oversight tools are attuned to the needs and preferences of all users involved regardless of their caregiving roles.



\textbf{\textit{Incentivizing Caregivers to Provide Oversight: }}
Caregivers may encounter a decrease in motivation under some circumstances, especially when they perceive fewer personal benefits or lack close relationships within the community \cite{akter_evaluating_2023, mendel_social_2023}. Therefore, it is important to incorporate reward mechanisms aimed at acknowledging and incentivizing their contributions as proactive digital citizens. Designers should also explore the development of effective nudging mechanisms strategically designed to prompt and encourage caregivers to engage more actively and consistently in providing oversight to their respective communities. For instance, weekly notifications or alerts upon new app installations within the community could prompt caregivers to assume oversight responsibilities. Implementing reward mechanisms, such as badges or points, to incentivize desired behaviors, could also be beneficial. Such interventions need to be carefully crafted to ensure that caregivees can also form a sense of duty and accountability to their communities.

\textbf{\textit{Enhancing Caregivees' Community Trust and Belonging: }}
Collaborative oversight should include mechanisms fostering caregivees' trust and community belonging. To do so, we need to explore ways to encourage caregivees to seek advice and guidance from their community so that they feel their privacy and security needs matter and they belong to the community. Implementing features like autogenerated text messages for advice can aid caregivees with effective assistance-seeking. It's also crucial to ensure these tools make caregivees feel comfortable and connected rather than surveilled or punished. Thoughtful design is essential to co-manage mobile privacy and security, safeguarding caregivees from potential misuse while enhancing their sense of community belonging.\\

\noindent
\textbf{Limitations and Future Work}\\
We recognize several limitations of our study that should be addressed in future work. First, our sample was skewed toward younger adults with at least bachelor's degrees. Our sample was also biased towards racial minorities with 72\% of our participants being of Asian (primarily South Asian) descent. Further, our caregivers and caregivees were also generally equally distributed across these ethnicities, age groups, and education levels. \edit{Moreover, the co-monitoring feature of CO-oPS may have introduced conflicts among caregivees, as indicated by our RQ3 results, which showed that caregivees hid significantly more apps than caregivers. This issue could be further exacerbated in communities with power hierarchies, such as parent-child relationships, or among ethnic groups with lower privacy boundaries.} Therefore, our results may not be generalizeable to populations of different ethnicities, age groups and education levels. Future work should explore caregiving with a community oversight mechanism among communities of more varied demographics and socio-economic status \cite{madden_privacy_2017}. 
One of the key limitations of our study was that we asked all of our participants to perform same set of tasks with CO-oPS, which resulted in non-significant differences in the behaviors of caregivees and caregivers. Additionally, the weekly tasks might have prompted caregivers and caregivees to adjust their app permissions or share oversight within their communities. In real-world scenarios, users might lack motivation for community oversight without incentives. Thus, we encourage further research into effective nudging mechanisms to incentivize users for community oversight and to take privacy protective actions for their own benefit. 

\edit{Another potential limitation is the risk of privacy breaches if a community member's phone is lost or stolen, as this could expose the personal information of the entire community. Additionally, in large communities that include members who may not be close, even though they were initially trusted, they could potentially become malicious. Future research should investigate the possibility of allowing more granular control over the app-hiding features, enabling community members to selectively turn off sharing with specific individuals. Furthermore, caregivers' advice and privacy behaviors may sometimes be misleading. In such cases, collaborative approaches need to identify deserving candidates to label as caregivers. Factors such as community members' power use, the validity of their advice, and their privacy practices could be important considerations for future studies.} \textcolor{black}{While we allowed all community members to hide any of their installed apps as they preferred, certain mobile apps carry different security and privacy risks \cite{reardon_50_2019, tandon_i_2022}. For example, caregivees were more likely to hide social media and shopping apps, which might have limited the feedback received for those apps. Future research should explore assigning risk-based weights to the apps installed, making users more aware of the risks associated with hiding certain apps and encouraging them to consider sharing to receive effective oversight.} Finally, in this study, we did not measure power use in the post study survey as we did not expect the CO-oPS app intervention to alter individual power usage. In retrospect, heightened self-efficacy might lead to increased power usage. Hence, we recommend that future studies on privacy interventions assess this factor both pre and post-intervention to gauge potential changes in power use.  
\section{Conclusion}
Collaborative mobile privacy and security approaches often treat all individuals in the community as equal, despite differences in expertise, needs and roles. Our study highlights the need to address these imbalances, as caregivers and caregivees have distinct perceptions and requirements. Our study illustrated the role community oversight can play in helping those who need support and build their self-efficacy and community collective-efficacy. Our findings call for interventions that prioritize increasing caregivees' community trust and belonging while incentivizing caregivers' participation in oversight. This approach aligns with our broader goal of developing community-based tools that effectively support mobile privacy and security for all. Future research will focus on refining these tools and exploring innovative strategies to enhance collaboration and address the unique needs of diverse community members.

\begin{acks}
We thank the individuals who participated in our study. We also acknowledge the contributions of Nazmus Sakib Miazi, Nikko Osaka, Anoosh Hari, Madiha Tabassum, and Ricardo Mangandi, in the CO-oPS app development. Our research was supported by the U.S. National Science Foundation under grants CNS-1814068, CNS-1814110, and CNS-2326901. Any opinion, findings, and conclusions or recommendations expressed in this material are those of the authors and do not necessarily reflect the views of the U.S. National Science Foundation.
\end{acks}
\bibliographystyle{ACM-Reference-Format}
\bibliography{main}


\newpage
 \onecolumn
 \begin{appendices}
\appendix


\section{Community Trust Survey Questionnaire}
Derived from Chouhan et al.'s conceptual model of Community Oversight \cite{chouhan_co-designing_2019}\\
1. I trust others in my community to protect my private information.\\
2. I trust others in my community to give me advice about mobile privacy and security.\\
3. Others in my community trust me to protect their private information.\\
4. Others in my community trust me to give them advice about mobile privacy and security.

\section{Sample Questions of Followup Interview}

  \textit{
    \begin{itemize}
\item During the study, how frequently did your community members discuss mobile privacy and security decisions with one another? 
\item During the study, how did you communicate with others who were part of your community? 
\item During the study, how did you manage your mobile privacy and security decisions? Did you see any changes compared to prior to the study? Why or why not?
\item Can you explain how and why the app did or did not help provide transparency into the mobile privacy and security decisions of other people in your community? 
\item How and why did the app or did not help raise awareness in your community about mobile privacy and security?
\item How and why did the app or did not enable you and individuals in your community to provide feedback and guidance about others’ mobile privacy and security?
\item How and why did the app or did not help you work together as a community about mobile privacy and security? 
\item Were there any problems or concerns you or others in your community encountered when using the app? 
\end{itemize}
} 

\section{Codebook}
\label{appendix:d}
 \begin{table*}[h]
 \centering
 \footnotesize
 \caption{Codebook of Interview Transcripts (28 Caregivers and 23 caregivees)}
   \label{tab:codebook}
\begin{tabular}{ |p{3.5cm}|p{10.8cm}|  }
\hline {\textbf{Themes}} & {\textbf{Illustrative Quotations}} \\ 
 \hline
\rowcolor{lightgray}
 \multicolumn{2}{|c|}{} \\[-8pt] 
 \multicolumn{2}{|c|}{\textbf{Reviewing Others' Apps}}  \\ \cline{1-2}

\textbf{Helped identify risky permissions} \newline 
Caregiver: 36\%; Caregivee: 74\%
 & \textit{“
 So one of my friends sent me a text to the co-ops app. This Prime bank has a lot of your app permissions, so please look at it. 
 And actually am not using that Prime bank account any more, but I forgot about this app and so I got to know this because he reviewed it, right.”} - C05P4, caregivee   \\ \cline{1-2}
   
 \textbf{Gave peace of mind}  \newline 
Caregiver: 29\%; Caregivee: 68\%
 & \textit{“I think having this app actually made the community, especially [Name], see these apps of our phones, it kinda give a peace of mind you know. Because now you know there is someone who keeps an eye on your security settings.”} - C02P1, caregivee   \\ \cline{1-2} 

 \textbf{ Distributed responsibility}  \newline
  Caregiver: 39\%; Caregivee: 9\%
 & \textit{“As checking all of someone's apps is not an easy task, right, you may miss some because there are so many apps people have. This app let others also check the apps. Thats a good thing. I think this is how the responsibility is shared. The more people check, the lower the workload.”} - C08P1, caregivee \\ \hline
\rowcolor{lightgray}
 \multicolumn{2}{|c|}{} \\[-8pt] 
 \multicolumn{2}{|c|}{\textbf{Sending Messages}}  \\ \cline{1-2}
   
 \textbf{Initiating discussion became easier} \newline
  Caregiver: 93\%; Caregivee: 43\%
 & \textit{“I am not close to any of the guys, but I think we can talk now with each other with the help of this app. 
 because, there are a lot of elements just for communicating, but this is kind of an app, we can easily talk about the app security things, even though we are not much close.”} - C06P2, caregiver \\ \cline{1-2}

\textbf{Seeking advice became easier}  \newline
Caregiver: 21\%; Caregivee: 83\%
& \textit{“
So whenever he installs any app, he used to go to settings to see the permissions of it. 
But the thing is that I didnt know, so now that we are using this app, and made me think about it and ask him about the privacy and security reasons and he instantly give me some opinion and some guidance.”} - C17P1, caregiver   \\ \cline{1-2}
   
\textbf{Keeping conversations for S\&P only} \newline  
Caregiver: 50\%; Caregivee: 17\%
 & \textit{“This was a very helpful feature for me, because whenever I wanted to say something to a particular person, I could directly use the app to send them a message... 
 This app allows to keep the conversation directly about app security. Otherwise, it would be awkward to have this conversation in the other messaging.”} - C05P3, caregiver    \\ \hline
\rowcolor{lightgray}
 \multicolumn{2}{|c|}{} \\[-8pt] 
 \multicolumn{2}{|c|}{\textbf{Changing Apps/Permissions}}  \\ \cline{1-2}

 \textbf{Comparisons led to change} \newline
Caregiver: 65\%; Caregivee: 100\%
 & \textit{“
 I saw that some of my friends didn't get permission to Facebook, on their location or something. So I see that in my app, I already give the permission of location. So after seeing that, 
 I removed my location permission from my Facebook. So that's how it helped.”} - C12P1, caregiver \\ \cline{1-2}

\textbf{Learned to change permissions} \newline
Caregiver: 39\%; Caregivee: 78\%
 & \textit{“
 But if I look at the apps list now and see all the permissions that are granted, I'm like, let me go to settings and change the permissions that makes sense for the app to get access to. So this changing of permissions are now a bit straightforward.”} - C18P4, caregiver   \\ \cline{1-2}
   
\textbf{Inspired from community posts}    \newline
Caregiver: 21\%; Caregivee: 61\%
 & \textit{“So there is like a community tab. So when they change app settings, it comes there. So even viewing from there, you can have more information that yeah, this app doesn't need that, let me check if my app uses that... I think in this case, it did actually allow me take the initiative.”} - C03P4, caregivee    \\ \hline
\rowcolor{lightgray}
 \multicolumn{2}{|c|}{} \\[-8pt] 
 \multicolumn{2}{|c|}{\textbf{Hiding Apps}}  \\ \cline{1-2}

\textbf{Affected personal privacy} \newline
Caregiver: 54\%; Caregivee: 100\%
 & \textit{“Another drawbacks is they would be able to know which apps I am using, or I would know their apps. These our personal apps, we probably would not like others to see. 
 ”} - C05P4, caregivee \\ \cline{1-2}

\textbf{Concerns for varied relationships} \newline
Caregiver: 32\%; Caregivee: 91\%
 & \textit{“you don't want your friends to be in the same community with your parents. Because you dont want your parents to see some of your apps or your friends to see some of your other apps, right? So yeah, I guess that was my main concern. … I also feel like I don't think my friends should be seeing my parents' apps, either.”} - C15P2, caregivee   \\ 
 
 \hline
 \textbf{Defeats the purpose} \newline
Caregiver: 39\%; Caregivee: 9\%
& \textit{“The problem is they can hide them. Of course, you want to have privacy, right? But here the purpose was to monitor your sons or your parents, if they're older, and then they can hide things, I think it sort of defeats the purpose.”} - C22P1, caregiver   \\ \hline
\end{tabular}
\end{table*}


\end{appendices}

\end{document}